\documentclass{emulateapj} 
\usepackage{graphicx}

\def\ltsima{$\; \buildrel < \over \sim \;$}
\def\lsim{\lower.5ex\hbox{\ltsima}}
\def\gtsima{$\; \buildrel > \over \sim \;$}
\def\gsim{\lower.5ex\hbox{\gtsima}}

\begin{document}

\title{The Effects of Irradiation on Hot Jovian Atmospheres: \\ Heat
Redistribution and Energy Dissipation}
\author{Rosalba Perna\altaffilmark{1}, Kevin Heng\altaffilmark{2,4} \& Fr\'{e}d\'{e}ric Pont\altaffilmark{3}}
\altaffiltext{1}{JILA and Department of Astrophysical and Planetary Sciences, 
University of Colorado, Boulder, CO 80309, U.S.A.}
\altaffiltext{2}{ETH Z\"{u}rich, Institute for Astronomy, Wolfgang-Pauli-Strasse 27, CH-8093, Z\"{u}rich, 
Switzerland}
\altaffiltext{3}{University of Exeter, Stocker Road, Exeter EX4 4QL, U.K.}
\altaffiltext{4}{Zwicky Prize Fellow}

\begin{abstract}
Hot Jupiters, due to the proximity to their parent stars, are
subjected to a strong irradiating flux which governs their radiative
and dynamical properties.  We compute a suite of 3D circulation models
with dual-band radiative transfer, exploring a relevant range of
irradiation temperatures, both with and without temperature inversions. We find that, for irradiation temperatures $T_{\rm irr}\lsim 2000$~K, heat redistribution is very
efficient, producing comparable day- and night-side fluxes. For
$T_{\rm irr}\approx 2200$--2400 K, the redistribution starts to break
down, resulting in a high day-night flux contrast.  Our simulations
indicate that the efficiency of redistribution is
primarily governed by the ratio of advective to radiative timescales.
Models with temperature inversions display
a higher day-night contrast due to the deposition of starlight at
higher altitudes, but we find this opacity-driven effect to be
secondary compared to the effects of irradiation.  The hotspot offset
from the substellar point is large when insolation is weak and
redistribution is efficient, and decreases as redistribution breaks
down. The atmospheric
flow can be potentially subjected to the Kelvin-Helmholtz instability
(as indicated by the Richardson number) only in the uppermost layers,
with a depth that penetrates down to pressures of a few millibars at
most. Shocks penetrate deeper, down to several bars in the hottest
model.  Ohmic dissipation
generally occurs down to deeper levels than shock dissipation (to tens
of bars), but the penetration depth varies with the atmospheric
opacity.  The total dissipated Ohmic power increases steeply with the
strength of the irradiating flux and the dissipation depth recedes
into the atmosphere, favoring radius inflation in the most irradiated
objects.  A survey of the existing data, as well as the inferences
made from them, reveals that our results are broadly consistent with
the observational trends.
\end{abstract}


\section{Introduction}

Hot Jupiters are a class of exoplanets believed to have their rotational
period synchronized with their orbital motion due to the strong tidal
forces resulting from the proximity to their parent stars.  Because of tidal locking, 
these exoplanets possess permanent day- and night-side hemispheres.  
These special conditions make hot Jupiters
exciting astrophysical laboratories, in which the interplay of
radiative forcing, atmospheric dynamics, composition, rotation speed
and surface gravity contribute to their observational properties.

A growing body of observations in recent years has allowed for hot Jovian atmospheres 
to be characterized via a combination of
different techniques, which include the measurements of transmission
spectra, thermal phase curves, and secondary eclipses.  A complex picture 
has emerged with the objects characterized by a wide
range of properties: some display temperature
inversions in their atmospheres, others do not; heat redistribution
is very efficient in some cases, but highly inefficient in others; a
fraction of exoplanets display radii much too large to be compatible with
standard evolutionary models.

Theoretical explanations for this variety of phenomena have been abundant.
The presence of temperature inversions in some exoplanets, but not in
others, has been shown to be an effect of atmospheric opacity (Fortney
et al. 2008).  The variety of recirculation efficiencies, on the other
hand, may be attributed to several mechanisms (e.g., Cowan et
al. 2012), such as shorter radiative times in hotter exoplanets, stronger
magnetic drag (which retards the wind speeds) at higher irradiation
fluxes, and a weaker greenhouse with increasing exoplanetary temperature.

For the problem of inflated radii, a variety of possible explanations
has been put forward (see, e.g., Fortney \& Nettelman 2010 for a recent
review). Bodenheimer et al. (2001) pointed out that tidal dissipation,
induced by a non-negligible orbital eccentricity, can pump energy into
the interior and hence counteract contraction during evolution.  This
idea has been further explored and refined by Jackson et al. (2008),
Miller et al. (2009) and Ibgui \& Burrows (2009).  Other ideas
(specifically proposed for the case of HD~209458b) include the tidal damping
of obliquity (Winn \& Holman 2005) and extreme evaporation (Baraffe
et al. 2004). Further, it was noted that enhanced opacities in the
atmosphere may retard cooling and hence delay contraction (Burrows
et et al., 2007a, but see also Guillot 2008).  \cite{gs02} have
suggested that a fraction of a percent or so of the absorbed stellar
flux can be converted into kinetic energy in the deep interior by the
breaking of atmospheric waves, but Burkert et al. (2005) were not able
to confirm this dissipation mechanism via numerical simulations.  More
recently, the observation that the flow of ionized particles in the
interior generates substantial atmospheric currents has led to
investigations which showed that Ohmic dissipation can inject enough
energy in the deep atmospheric levels to affect the evolution of the
radius (Batygin \& Stevenson 2010; Perna et al 2010b).

Ultimately, discrimination among various models will happen through
a close comparison with the data, as more accumulate.
From an observational standpoint alone, current data strongly hint at
the fact that {\em the strength of the irradiating flux} plays a major role
in determining the degree of heat redistribution (Harrington  2011), as well as the
amount of radius inflation (e.g., Fortney et al. 2010). This is unsurprising, 
since irradiation is the main energy provider in these exoplanets. 
However, from a theoretical point of view, the direct link  between strength
of irradiation and those fundamental observational properties of the
exoplanets has not been elucidated yet. This has motivated our current work.

In this paper, by means of global, 3D atmospheric circulation models with dual-band
radiative transfer tailored to hot Jupiters \citep{hfp11}, we explore,
as a function of the irradiating flux from the parent star: \\ {\em
  a)} The efficiency of heat redistribution from day to night side,
focusing on the infrared photosphere, and the relative day-night brightness
contrast;\\ {\em b)} The offset of the hotspot (hottest infrared region of
the exoplanet) with respect to the substellar point;\\ {\em c)} The
magnitude of hydrodynamic dissipation due to both the onset of the
Kelvin-Helmholtz instability and shocks;\\ {\rm d)} The strength of
Ohmic dissipation.\\ We present a suite of 16 models: for eight values
of the irradiation temperature (spanning the range $T_{\rm irr}
\approx 770-3000$~K), and each for cases with and without a temperature
inversion.

Our simulations allow us to elucidate the relative importance of
irradiation and opacity (correlated with temperature inversions) in
determining the observational properties of the exoplanet. They also allow us to assess
what is the main effect in determining heat redistribution and
day-night contrast, and to predict theoretically, for the first time,
the temperature or irradiating flux around which redistribution begins to break
down. Furthermore, by computing the Richardson number and the Mach
number at each point in the flow, we can quantitatively estimate
the depth down to which the flow can be subject to either shear
instabilities or shocks, and hence whether hydrodynamic dissipation
of this type can affect radius inflation. Similarly, our simulations
allow us to follow the depth down to which Ohmic power can be
dissipated, as a function of the irradiation strength, and for
models with varying opacities.

Our paper is organized as follows: in \S2, we summarize  the main
features of our  simulations, and provide the specific  details for the
models that we study. Our results for heat redistribution and the hotspot
location are presented in \S3, while in \S4 we study hydrodynamic
and Ohmic dissipation. A qualitative comparison with observations is made in \S5.
We finally summarize and conclude in \S6.  For completeness, we provide zonal-mean profiles of the zonal wind, temperature, potential temperature and streamfunction in Appendix \ref{append}.

\section{Methodology}
\label{sect:method}

Our 3D atmospheric circulation models are based on the computational setup described in \cite{hmp11} and \cite{hfp11}. 
In the following, we summarize its main features, as well as the set of parameters
which we use for the current simulations.

\begin{table*}
\centering
\caption{Table of common parameters and their values}
\label{tab:params}
{\footnotesize
\begin{tabular}{lccc}
\hline\hline
\multicolumn{1}{c}{Symbol} & \multicolumn{1}{c}{Description}  & \multicolumn{1}{c}{Units} & \multicolumn{1}{c}{Value(s)}\\
\hline
\vspace{2pt}
$\tau_{\rm S_0}$ & normalization for optical depth of shortwave absorbers & --- & $5 \times 10^3, 5 \times 10^4$ \\
$\tau_{\rm L_0}$ & normalization for optical depth of longwave absorbers & --- & $10^4$ \\
$\epsilon$ & enhancement factor due to collision-induced absorption & --- & 1 \\
$c_P$ & specific heat capacity at constant pressure (of the atmosphere) & J K$^{-1}$ kg$^{-1}$ & 14550.4 \\
${\cal R}$ & specific gas constant (of the atmosphere) & J K$^{-1}$ kg$^{-1}$ & 4157.25 \\
$\kappa \equiv {\cal R}/c_P$ & adiabatic coefficient$^\ddagger$ & J K$^{-1}$ kg$^{-1}$ & 2/7 \\
$P_0$ & reference pressure at bottom of simulation domain & bar & 1000 \\
$g_p$ & acceleration due to gravity & m s$^{-2}$ & 10 \\
$R_p$ & radius of hot Jupiter & km & $7.1 \times 10^4$ \\
$t_\nu$ & hyperviscous time & day$^\dagger$ &$10^{-6}$ \\
$N_{\rm v}$ & vertical resolution & --- & 36 \\
\hline
\hline
\end{tabular}}\\
$\dagger$: expressed in terms of an exoplanetary day (i.e., rotational period).\\
$\ddagger$: we term $\kappa$ the ``adiabatic coefficient", following \cite{p10}, and avoid calling it \\
the ``adiabatic index" in order to not confuse it with the ratio of specific heat capacities.
\end{table*}

\subsection{Stellar and orbital parameters}

A concise list of the parameters and their adopted values are
described in Table \ref{tab:params}.  Most of the parameters are
chosen to have typical values, e.g., the surface gravity of the hot
Jupiter is $g_p = 10$ m s$^{-2}$. Since our primary interest is to explore the 
dependence of selected exoplanetary properties on the strength of irradiation 
and the shortwave atmospheric opacity, we keep the rest of the parameter 
values to be the same for all of the models (Table \ref{tab:params}).

Assuming a Bond albedo of zero, the irradiation\footnote{In this study, we use the terms ``irradiation" and ``isolation" interchangeably.} temperature is defined as \citep{hhps12}
\begin{equation}
T_{\rm irr} = T_\star \left( \frac{R_\star}{a} \right)^{1/2},
\end{equation}
such that the irradiating flux is ${\cal F}_0 = \sigma_{\rm SB}
T^4_{\rm irr}$.  (For parameter values appropriate to Earth, one
recovers ${\cal F}_0 \approx 1370$ W m$^{-2}$, the solar constant.)
The irradiating flux ${\cal F}_0$, stellar radii $R_\star$, effective stellar temperature $T_\star$ and exoplanet-star spatial
separation $a$ are related by the expression,
\begin{eqnarray}
a &=& R_\star T^2_\star \left( \frac{\sigma_{\rm SB}}{{\cal F}_0}
\right)^{1/2}
\approx 0.09 \mbox{ AU} \left( \frac{R_\star}{R_\odot}
\right) \nonumber \\ &\times &
\left( \frac{T_\star}{6000 \mbox{ K}} \right)^2
\left( \frac{{\cal F}_0}{2 \times 10^8 \mbox{ erg cm}^{-2} \mbox{ s}^{-1}} \right)^{-1/2},
\end{eqnarray}
where $\sigma_{\rm SB}$ is the Stefan-Boltzmann constant and we have
assumed that the exoplanet orbits a Sun-like star.  The use of the irradiation, rather than 
the equilibrium temperature, circumvents the issue of dealing with (confusing) factors of order unity 
related to assumptions about the efficiency of heat redistribution.

The tidal locking time is \citep{gs66,bod01}
\begin{equation}
t_{\rm lock} \approx \frac{8Q}{45 \Omega_p} \left( \frac{\omega_p}{\Omega_p} \right) \left( \frac{M_p}{M_\star} \right) \left( \frac{a}{R_p} \right)^3,
\end{equation}
where $\omega_p$ and $\Omega_p$ are the rotational and orbital frequencies of the exoplanet, respectively.  For our models, we get $t_{\rm lock} \sim 10^3$--$10^8$ yr $(Q/10^6) (\omega_p/\Omega_p)$.
Thus, for a Gyr-old solar-like star (which is what we assume), our
assumption of tidal locking is plausible and self-consistent if $\omega_p/\Omega_p \sim 1$.  If $\omega_p/\Omega_p \sim 10$, then this assumption becomes suspect only for Model C.  With the rotational and orbital frequencies being equal, we obtain
\begin{equation}
\Omega_p \approx \left(\frac{G M_\star}{a^{3}}\right)^{1/2},
\end{equation}
where $G$ is the gravitational constant and $M_\star$ is the stellar
mass.  Thus, ${\cal F}_0$, $\Omega_p$ and $a$ need to be varied
self-consistently.
We note that setting the rotational and orbital frequencies to be
  equal results in a more constrained problem for our simulations.
  While the existence of close-in, non-spin-synchronized gas giants
  remains possible, there is little empirical evidence currently
  available to support such a modelling effort.

We present 16 models in this paper, divided into subgroups which we
term ``cold" (C, C1 and C2), ``warm" (W, W1 and W2) and ``hot" (H,
H1).  For Model W, the irradiating flux ${\cal F}_0$ is chosen to
match the threshold value found by \cite{ds11}, below which transiting
\textit{Kepler} giant exoplanet candidates appear to have non-inflated
radii. Each of these models is studied both with and without the
presence of a temperature inversion.  The parameters specific to each
model are reported in Table~\ref{tab:params2}.


\begin{table*}
\centering
\caption{Table of parameters, and their values, specific to each model}
\label{tab:params2}
{\scriptsize
\begin{tabular}{lcccccccc}
\hline\hline
\multicolumn{1}{c}{Model} & \multicolumn{1}{c}{${\cal F}_0$ (W m$^{-2}$)}  & \multicolumn{1}{c}
{$\Omega_p$ (s$^{-1}$)} & 
\multicolumn{1}{c}{a (AU)} & \multicolumn{1}{c}{$T_{\rm init}$ (K)} & \multicolumn{1}{c}{$\Delta t$ (s)} 
& \multicolumn{1}{c}{Simulation time (Earth days)} & $C_{\rm int}$ (J K$^{-1}$ m$^{-1}$) & $t_\nu^{-1}$ (s$^{-1}$) \\
\hline
C & $2 \times 10^4$ & $1.3 \times 10^{-6}$ & 0.28 & 589, 501 & 600 & 4500 & $10^7$ & $2.1 \times 10^{-1}$ \\
C1 & $5 \times 10^4$ & $2.6 \times 10^{-6}$ & 0.17 & 741, 630 & 240 &  3000 & $10^7$ & $4.1 \times 10^{-1}$ \\
C2 & $7 \times 10^4$ & $3.4 \times 10^{-6}$ & 0.15 & 806, 686 & 240 & 3000 & $10^7$ & $5.4 \times 10^{-1}$ \\
W & $2 \times 10^5$ & $7.5 \times 10^{-6}$ & 0.09 & 1048, 892 & 180 & 2000 & $10^6$ & 1.2 \\
W1 & $5 \times 10^5$ & $1.5 \times 10^{-5}$ & 0.06 & 1317, 1121 &  120 & 1500 & $10^6$ & 2.4 \\
W2 & $7 \times 10^5$ & $1.9 \times 10^{-5}$ & 0.05 & 1433, 1219 & 120 & 1500 & $10^6$ & 3.0 \\
H & $2 \times 10^6$ & $4.2 \times 10^{-5}$ & 0.03 & 1863, 1585 & 120 & 1500 & $10^6$ & 6.7 \\ 
H1 & $5 \times 10^6$ & $8.4 \times 10^{-5}$ & 0.02 & 2343, 1994 & 80 & 1000\ & $10^6$ & 13.3 \\
\hline
\end{tabular}}\\
Note: initial temperatures quoted are for $\gamma_0=0.5$ and 2.0, respectively.
\end{table*}

\subsection{Thermodynamics}

The adiabatic coefficient $\kappa$ is related to the thermodynamical properties of the atmospheric gas \citep{p10},
\begin{equation}
\kappa \equiv \frac{{\cal R}}{c_P} = \frac{2}{2+n_{\rm dof}}.
\end{equation}
By assuming the atmosphere to be dominated by molecular hydrogen
(i.e., the number of degrees of freedom of the gas is $n_{\rm
  dof}=5$), we get $\kappa=2/7$.  Furthermore, the specific gas
constant is related to the mean molecular weight $\mu$
and the universal gas constant ${\cal R}^\ast$ by ${\cal R} =
{\cal R}^\ast/\mu = 4157.25 $ J K$^{-1}$ kg$^{-1}$.  
We use $\mu=2$ to emphasize the detachment from any specific case
  study, and remark that our results are rather insensitive
  to small changes in the mean molecular weight (e.g., $\mu=2.35$
as is typical for a solar mix).  It follows that the
specific heat at constant pressure is $c_P = 14550.4$ J K$^{-1}$
kg$^{-1}$.  We recall that the adiabatic index is $\gamma = 1 +
2/n_{\rm dof} = 7/5$ for an atmosphere dominated by molecular hydrogen
and that the sound speed is $c_s = \sqrt{\gamma k_{\rm B} T/ 2 m_{\rm
    H}}$ with $k_{\rm B}$ denoting the Boltzmann constant, $T$ the
temperature and $m_{\rm H}$ the mass of a hydrogen atom.

\subsection{Atmospheric opacities}

In this work, we adopt the dual-band approximation for radiative
transfer, namely that the blackbody peaks of the stellar and
exoplanetary emission are well separated in wavelength or frequency
(see \citealt{hfp11} and references therein).  One then requires two
broadband opacities to completely describe the radiative transfer: a
shortwave/optical opacity $\kappa_{\rm S}$ to quantify the absorption
of starlight with depth/pressure and a longwave/infrared opacity
$\kappa_{\rm L}$ to quantify the absorption of thermal emission from
the exoplanet.  We neglect the effects of scattering (see \citealt{heng12}).  Varying
$\kappa_{\rm S}$, while holding $\kappa_{\rm L}$ fixed, mimics the
effect of extra, shortwave absorbers in the atmosphere, of unspecified
chemistry, which cause a temperature inversion when $\kappa_{\rm S} >
\kappa_{\rm L}$ (Hubeny, Burrows, \& Sudarksy 2003, Burrows et
al. 2007b, Burrows \& Orton 2010, Knutson et al. 2008).

\begin{figure}
\centering
\includegraphics[width=9cm]{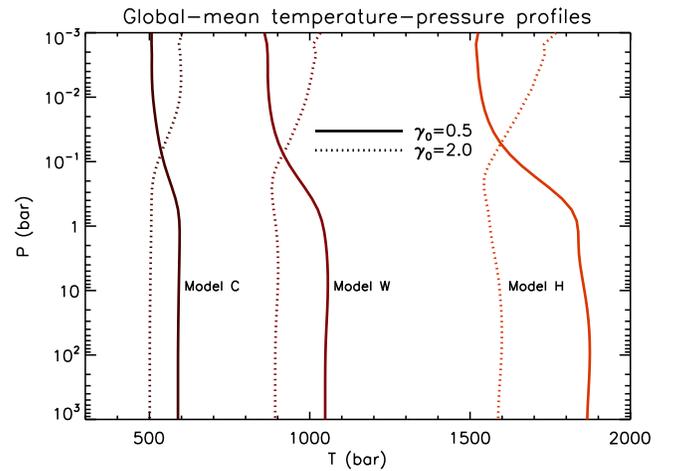}
\caption{The global-mean temperature-pressure profiles for Models C, W and H.
Models with $\gamma_0<1$ do not display temperature inversions, while models with
$\gamma_0>1$ do.}
\label{fig:tp}
\end{figure}

In the absence of empirical constraints, we assume the shortwave
opacity to be constant, which means that it is related to the
shortwave optical depth in a simple manner,
\begin{equation}
\tau_{\rm S} = \frac{\kappa_{\rm S} P}{g_p} = \frac{\tau_{\rm S_0} P}{P_0},
\end{equation}
where $P_0 = 1$ kbar is the reference pressure at the bottom of the
simulation domain.  The longwave optical depth is described by two
terms \citep{hfp11},
\begin{equation}
\tau_{\rm L} = \tau_{\rm L_0} \left[ \frac{P}{P_0} + \left( \epsilon -1 \right) \left( \frac{P}{P_0} \right)^2 \right],
\end{equation}
where the second term is an approximate way to account for the effects
of collision-induced absorption, which introduces an enhancement
factor of $\epsilon$ at the bottom of the simulation domain.  The
longwave optical depth normalization is $\tau_{\rm L_0} = \kappa_0
P_0/g_p$, where $\kappa_0$ is the longwave opacity normalization.  We pick $\epsilon = 1$.
Ignoring collision-induced absorption in the infrared is a good
approximation as long as this effect is only significant at
atmospheric layers residing below the longwave photosphere (Heng, Frierson \& Phillipps 2011).

In the absence of clouds/hazes, the ratio $\gamma_0 \equiv \kappa_{\rm S}/\kappa_0$ 
determines if a temperature inversion exists in the
atmosphere (Hubeny, Burrows, \& Sudarksy 2003; see also discussion in \citealt{hhps12}).
If $\gamma_0 > 1$, the shortwave photosphere sits above the
longwave photosphere, and an inversion exists; if
$\gamma_0 < 1$, the relative locations of the two photospheres are reversed, and
there is no inversion.  To study the effect of a temperature
inversion on the day-night heat redistribution, we examine models with
$\gamma_0 = 0.5$ and 2.  We choose a typical value for the longwave
opacity normalization: $\kappa_0 = 0.01$ cm$^2$ g$^{-1}$ such that the longwave/infrared 
photosphere resides at 
\begin{equation}
P_{\rm IR} \sim \frac{g_p}{\kappa_0} = 0.1 \mbox{ bar}.
\end{equation}
The shortwave opacity follows from the chosen value of $\gamma_0$.  

The global-mean temperature-pressure profiles are showed in
Fig.~\ref{fig:tp}, for the two chosen values of $\gamma_0 = 0.5$ and
2, and the three representative flux strengths indicated with models
C, W, H in Table \ref{tab:params2}.  The effect of the temperature inversion
is clearly evident: models with $\gamma_0 = 0.5$ are hotter in the
interior, and cooler in the outerparts, while models with $\gamma_0 =
2$ display the opposite behaviour.  We have set the internal heat flux 
of our hot Jupiter models to be zero.  For completeness, we include in Appendix \ref{append} the zonal-mean zonal wind, temperature, potential temperature and streamfunction profiles corresponding to these 6 models.

\begin{figure}
\centering
\includegraphics[width=9cm]{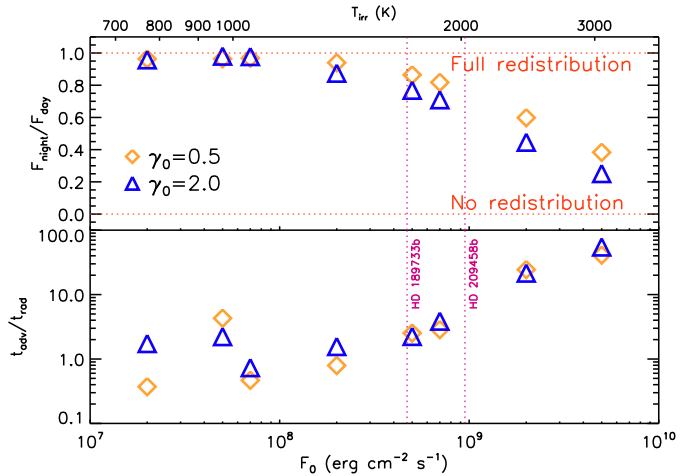}
\caption{{\em Top: } Ratio of night- to day-side hemispheric, infrared fluxes, for models
with various degrees of irradiation. The night-day contrast decreases with $T_{\rm irr}$,
indicating a weakening of the heat redistribution with increasing irradiating fluxes. Models
with temperature inversions display a larger flux contrast, but this is a secondary effect compared to the irradiating flux.
{\em Bottom:} Ratios between the advective and the radiative timescale, for the models
of the top panel. Redistribution begins to break down when the advective timescale
becomes much larger than the radiative time.}
\vspace{0.1in}
\label{fig:contrast}
\end{figure}

\subsection{Atmospheric dynamics}

The Princeton-GFDL \texttt{FMS} dynamical core solves the primitive
equations of meteorology (see \citealt{hmp11} and references therein).
In the vertical, we choose the simulation domain to span 6 orders of
magnitude in pressure, from $P=1$ mbar to 1 kbar; the latter value is
chosen to roughly match the transition from the radiative to the
convective zone of a hot Jupiter.  The vertical resolution is $N_{\rm
  v}=36$, such that each pressure scale height is covered by about 3
grid points.  The horizontal resolution is $192 \times 96$ (longitude
versus latitude).  In the absence of a physically-motivated scheme for
more sophisticated initial conditions, the simulations are started
from a state of rest with a constant temperature $T_{\rm
  init}$.\footnote{This particular initial condition is
  inconsequential for our simulations, since the isothermal initial
  condition in the code reproduces the radiative solution \citep{hfp11}.}  For $\epsilon=1$, the initial temperature is \citep{hfp11}
\begin{equation}
T_{\rm init} \approx \left[ \frac{{\cal F}_0}{16 \sigma_{\rm SB}} 
\left( 2 + \frac{\sqrt{3}}{\gamma_0} \right) \right]^{1/4}.
\end{equation}
The specific value of $T_{\rm init}$ used to initiate each of the models
is reported in Table~\ref{tab:params2}.

The areal heat capacity of the simulation bottom ($C_{\rm int}$) is
chosen to have a value such that the thermal inertia of gaseous
hydrogen matches the radiative time constant at $P=1$ kbar
\citep{hfp11}.  For Models C, W and H, it suffices to adopt $C_{\rm
  int} = 10^7, 10^6$ and $10^6$ J K$^{-1}$ m$^{-2}$, respectively, as
it has been shown by \cite{hfp11} that order-of-magnitude variations
in $C_{\rm int}$ produce small ($<1$ K) variations in the
temperature-pressure profile.

As already noted by \cite{hmp11}, numerical noise accumulates at the
grid scale and has to be dissipated via a ``hyperviscosity" in our
spectral code.  Since the hyperviscous term is neither an original nor a
physical part of the primitive equations, its value cannot be
specified from first principles.  As in \cite{hmp11} and \cite{hfp11},
we pick the order-of-magnitude value of the hyperviscous timescale to
be as large as possible in order for the simulations to collectively
run to completion: $t_\nu = 2\pi \times 10^{-6}/\Omega_p$.  Decreasing
the hyperviscous timescale --- and hence \emph{increasing} the rate of
dissipation ($t_\nu^{-1}$) --- by one to two orders of magnitude
leaves the hotspot offset largely unchanged (see Fig. 19 of
\citealt{hfp11}), but decreases the zonal wind speeds by $\sim 10\%$
\citep{hmp11}.  Our estimates for the zonal wind speeds are thus upper
limits with regards to this numerical issue.

Finally, we note that convective adjustment is applied when the dry lapse rate becomes super-adiabatic, i.e., the Schwarzschild criterion for convective stability is violated.  Our convective adjustment scheme is described in \cite{hfp11}.

\begin{figure}
\centering
\includegraphics[width=9cm]{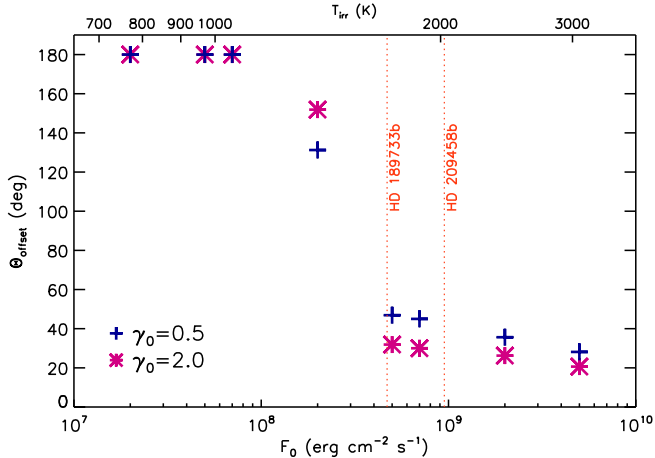}
\caption{Longitudinal separation between the location of the brightest spot 
on the exoplanet and the substellar point, i.e., the hotspot offset. Note that in the coldest models, the day-night contrast is negligible, and the location of the brightest spot
is subject to statistical fluctuations with time. When these conditions occur,
we set the offset equal to 180$^\circ$.}
\vspace{0.1in}
\label{fig:offset}
\end{figure}

\section{Atmospheric circulation, heat redistribution, and hotspot location}

In a strongly irradiated exoplanet, the atmospheric dynamics and radiation are regulated by the interplay of zonal winds, which tend to redistribute/transport heat to cooler regions
of the exoplanet, and local re-radiation of the absorbed energy, which favors the emission of radiation at the same location where it was absorbed. The relative efficiency of these two processes can be estimated by means of the associated timescales:
\begin{equation}
t_{\rm adv}\;\sim\; \frac{R}{{\mbox{max}\{ v_\Theta} \} }\,,
\label{eq:tau-adv}
\end{equation}
represents the fastest timescale on which zonal winds redistribute temperature variations
over the scale of the exoplanet, while the radiative timescale \citep{gy89}
\begin{equation}
t_{\rm rad}\;\sim\; \frac{c_P P}{\sigma_{\rm SB} g_p T_{\rm max}^3}
\label{eq:tau-rad}
\end{equation}
is the characteristic timescale on which thermal energy is radiated away by a parcel of fluid.  The latter is given by the ratio of the thermal heat content ($c_P T_{\rm max} \tilde{m}$ where $\tilde{m} = P/g_p$ is the column mass in hydrostatic equilibrium) to the re-emitted flux ($\sigma_{\rm SB} T_{\rm max}^4$).  Note 
that $T_{\rm max} \equiv T( \mbox{max}\{v_\Theta\} )$.  In other words, both timescales are evaluated at the 
location where the zonal velocity attains its maximum value.  Equation (\ref{eq:tau-rad}) is valid at optical depths close to or greater than unity; more precise versions of it carry an order of unity correction factor related to the efficiency of heat redistribution.

Hot Jovian atmospheres possess radiative timescales which vary across
many orders of magnitude.  At the top of the atmosphere ($\sim 1$
mbar), the atmosphere is radiative ($t_{\rm rad} \ll t_{\rm adv}$); at
the bottom ($\sim 1$ kbar), it is advective ($t_{\rm rad} \gg t_{\rm
  adv}$). The shortwave opacity controls the height/depth at which the
bulk of the starlight is deposited.  If the deposition
occurs mostly near the top of the atmosphere (high $\kappa_{\rm S}$),
then the hotspot offset is small.  Conversely, if it occurs near the
bottom, then the offset is large.  Thus, there is a degeneracy of the
hotspot offset on irradiation and optical opacity.

We denote the flux emerging from the longwave photosphere by ${\cal F}_{\rm OLR}$ (where ``OLR" denotes the ``outgoing longwave radiation", a term commonly used in the atmospheric/climate science community).
Denoting the latitude by $\Phi$, its latitudinal average is \citep{ca08}
\begin{equation}
\langle {\cal F}_{\rm OLR} \rangle \equiv \frac{1}{\pi} \int^{\pi/2}_{-\pi/2} ~{\cal F}_{\rm OLR} ~\cos^2\Phi ~d\Phi\,.
\end{equation}
With the longitude denoted by $\Theta$, the total flux from each hemisphere of the exoplanet is defined as
\begin{equation}
{\cal F}_{\rm day,night}\;=\;\frac{1}{\pi}\int_{\Theta_{\rm min}}^{\Theta_{\rm max}}
\langle {\cal F}_{\rm OLR}\rangle \, d\Theta\,,
\label{eq: fluxtot}
\end{equation}
where $\left\{\Theta_{\rm min}, \Theta_{\rm max} \right\}
=\left\{ \pi/2, 3\pi/2 \right\}$ for the day side.  The night side is described by 
$\left\{\Theta_{\rm min}, \Theta_{\rm max} \right\}
=\left\{ 0, \pi/2 \right\}$ and $\left\{ 3\pi/2, 2\pi \right\}$.  
The substellar point is set at $\Theta=\pi$.  In computing the day-night flux contrast, we 
have chosen to deal entirely with the latitudinally-averaged OLR or 1D ``brightness map", rather than
the thermal phase curve which is a convolution of the physical distribution of flux across longitude and 
its geometric projection to the observer \citep{ca08}.  Additionally, we choose to plot the ratio of night-to-day fluxes, 
rather than the more familiar day-to-night ratio, since this is a quantity which ranges between 0 and 1 
(rather than between 1 and infinity).  However, our discussion will be mostly in the context of day-to-night flux ratios, 
which we generally refer to as the ``day-night contrast".

Figure \ref{fig:contrast} shows the day-night contrast 
for the 16 reference models that we have studied.
The three coldest models C, C1 and C2 display a near perfect redistribution of the heating, with
comparable day and night fluxes. A slight contrast starts to appear 
for the W model (irradiated by ${\cal F}_0=2\times 10^8$~erg~cm$^{-2}$~s$^{-1}$), and it
then gradually increases with the strength of the irradiating flux.  For the same
value of ${\cal F}_0$, models with temperature inversions display a larger day-night flux contrast since the bulk of the starlight is deposited higher up in the atmosphere where the flow is dominated by radiative cooling rather than advection.
The dependence of the flux contrast on opacity (and hence on the presence
or not of temperature inversions) is consistent with the results of
Fortney et al. (2008) using 1D models (see also Dobbs-Dixon \& Lin 2008).

The bottom panel of Fig.~\ref{fig:contrast} shows the ratio between the
advective and the radiative timescales as previously defined, for each of the models of the upper panel.
Recall that, since flow velocities vary with latitude, and so do the physical
distances traveled by the flow from one side to the other of the exoplanet,
the advective time is a function of $\Phi$. To assign a unique value
in Fig.~\ref{fig:contrast}, we searched for the latitude at which the
zonal flow attains its maximum speed, and computed the timescales at that 
latitude.  For the coldest cases, the advective and the radiative
times are within an order of magnitude of each other.  The fluctuations in $t_{\rm adv}/t_{\rm rad}$ arise from the fact that the difference between $T_{\rm max}$ and the mean temperature of the flow becomes less pronounced as the strength of irradiation decreases.  In other words, describing whether the zonal flow is predominantly advective or radiative by a single number becomes a less robust exercise when the flow becomes more zonally symmetric.  Instead, the day-night flux contrast is a better indicator.

As the irradiating flux becomes more intense, the ratio $t_{\rm adv}/t_{\rm
  rad}$ rapidly increases. Our numerical results hence support the
qualitative expectation that it is the relative magnitude of these
two timescales which mainly determines the extent of heat redistribution, 
and consequently the day-night flux contrast \citep{sg02}.  
We note that, while our study has
focused on spin-synchronized planets, the qualitative considerations
derived above are expected to hold more generally. A change in the
planet spin for the same $T_{\rm irr}$ would modify the flow dynamics
(and hence both $t_{\rm rad}$ and $t_{\rm adv}$); however, the extent
to which the flow is able to redistribute heat versus dissipate it
locally, would still depend on the relative magnitude of these
timescales, as long as the spin period is not  much smaller
than $t_{\rm rad}$ (in which case the planet would be kept at a
uniform temperature).

The longitudinal offset $\Theta_{\rm offset}$ between the hotspot
location and the substellar point is displayed in Fig.~\ref{fig:offset}. 
For the three coldest models, the temperature is virtually homogeneous
on the entire exoplanet, but is subjected to statistical fluctuations; 
hence the location of the hottest spot is not well-defined. For these
cases, we set $\Theta_{\rm offset}=180^\circ$. As the day-night
contrast starts to deviate from unity, the hotspot location becomes
well-defined. In the W model, the hotspot offset from the substellar point
is rather large, $\Theta_{\rm offset}\approx 130^\circ$--$150^\circ$; this is
indicative of efficient heat redistribution. 
As the irradiating flux increases, the hotspot location moves towards
the substellar point. This can be easily understood in light of the 
changing ratio between the advective and the radiative timescales. As advection
becomes less effective in keeping up with the rapid cooling of the gas,
the hotspot location moves closer to the point of maximum irradiation, i.e., the 
substellar point.    

We note that the effects of opacity and mean molecular weight were
explored by \cite{menou12a} using also an atmospheric circulation code
with dual-band radiative transfer \citep{rm12}, albeit in the context
of Neptune-like exoplanets.  Our results are qualitatively consistent
with those reported in \cite{menou12a}, where it was found that
variations in the shortwave opacity produce non-negligible differences
in the hotspot offset.  Other studies of the role of atmospheric
  opacity in controlling the day/night contrast and dynamics have been
  performed by Lewis et al. (2010), tailored to the hot Neptune
  GJ436b. Their simulations particularly illustrated the important
  role of metallicity in regulating atmospheric circulation for a
  broad range of warm extrasolar planets.

\section{Dissipation within the atmospheric flow}

In this section, we use the 16 models previously described to
investigate the importance of various forms of dissipation within the
flow, and their relative trends with the strength of the irradiating
flux.  In particular, we first estimate the contribution to
hydrodynamic dissipation from gas which is either Kelvin-Helmholtz
unstable and/or shocked, and then we study the additional contribution
from Ohmic dissipation if the flow is magnetized.

We remark that, since our simulations do not self-consistently account
for the feedback effect of shocks, instabilities, and magnetic drag
(all contributing to reducing wind speeds and hence the amount of
dissipation), the estimates made in this section should be considered
as upper limits for each particular combination of
parameters.  However, the trends that we find, while likely to be
less steep with the presence of feedback effects, are expected to be
robust with respect to the dependence on irradiation.

\subsection{Hydrodynamic dissipation}

Our 3D circulation models, while not directly simulating the onset and development of shocks
and shear instabilities, allow us to establish whether the flow is
subject to such instabilities and to shocks and to determine the size of the
affected region. In the current study, we are primarily interested in exploring the 
effects of irradiation, while also examining the influence of (shortwave) opacity.

\begin{figure}
\centering
\includegraphics[width=9cm]{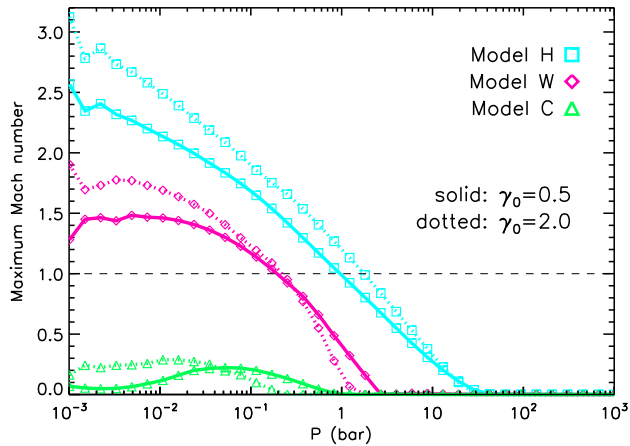}
\caption{Maximum Mach number as a function of depth within the atmosphere, for
models with varying degrees of irradiation.}
\vspace{0.1in}
\label{fig:mach}
\end{figure}

\begin{figure}
\centering
\includegraphics[width=9cm]{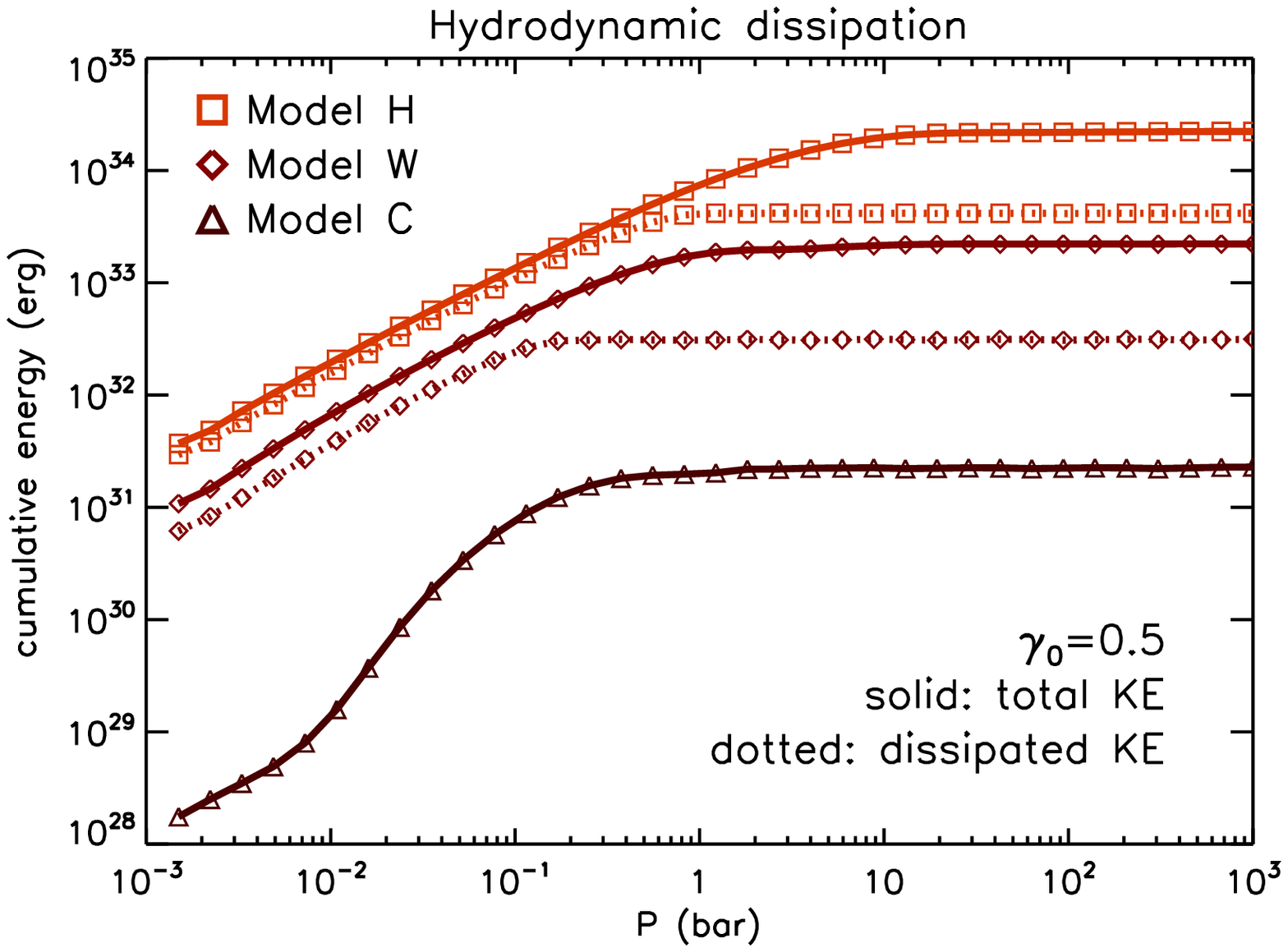}\\
\includegraphics[width=9cm]{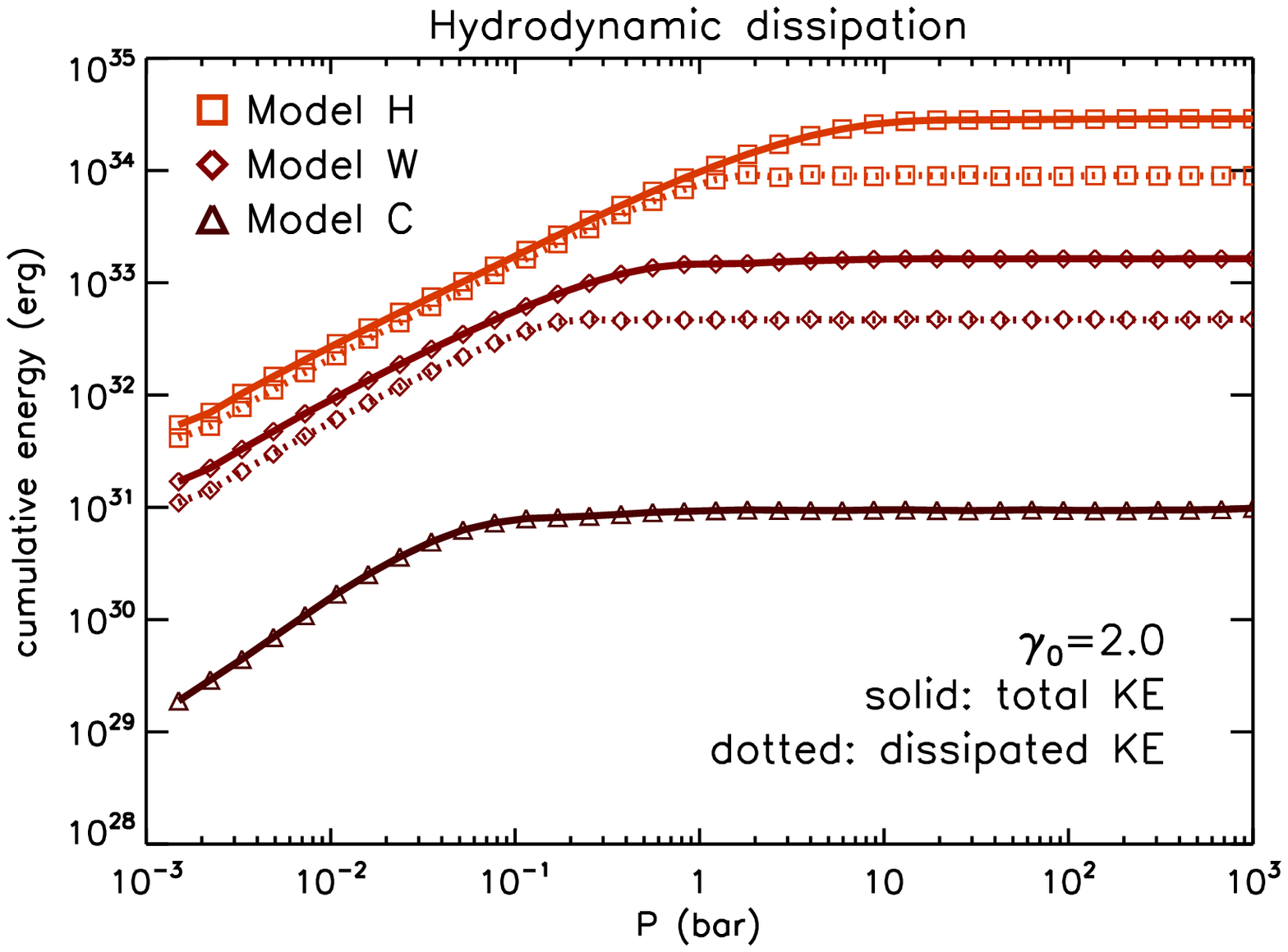}
\caption{{\em Top:} Kinetic energy available within the flow, together with the amount
available for dissipation through shear instabilities and shocks, for the three
representative models C, W, H  and no temperature inversions.
{\em Bottom:} Same as above, but for the C, W, H models with temperature inversions.}
\vspace{0.08in}
\label{fig:energy}
\end{figure}

\subsubsection{Dissipation from Kelvin-Helmholtz-unstable gas}

The first form of hydrodynamic dissipation that we consider is due to 
the Kelvin-Helmholtz (KH) instability in the thermally-stratified shear flow. 
Whether such a flow becomes unstable can be estimated by measurements
of the Richardson number
\begin{equation}
{\cal R}i = N^2 \left( \frac{\partial v_\Theta}{\partial z}
\right)^{-2} = -{\cal R} \sigma_0^{\kappa-1} \frac{\partial
  \theta_T}{\partial \sigma_0} \left( \frac{\partial
  v_\Theta}{\partial \sigma_0} \right)^{-2},
\label{eq:Ri}
\end{equation}
where
\begin{equation}
N = \left( \frac{g_p}{\theta_T} \frac{\partial \theta_T}{\partial z} \right)^{1/2}
\end{equation}
is the Brunt-V\"ais\"al\"a frequency, $\theta_T\equiv T(P/P_0)^{-\kappa}$ is
the potential temperature, and $\sigma_0 \equiv P/P_0$.
The criterion for the flow to become Kelvin-Helmholtz unstable is 
expressed by the condition ${\cal R}i<1/4$ \citep{kundu04,lg10}.

We computed the Richardson number at each point of the flow.  For Model C, 
we find ${\cal R}i>1/4$ everywhere, and hence shear
instabilities are not expected at low levels of irradiation. In the
warm and hot models, we find ${\cal R}i<1/4$ only in the outermost
pressure levels, down to a few millibars.  Hence, we conclude from this analysis
that the onset and consequent dissipation of shear instabilities is
unlikely to play a major role in inflating hot Jupiters.

\begin{figure*}
\centering
\includegraphics[width=18cm]{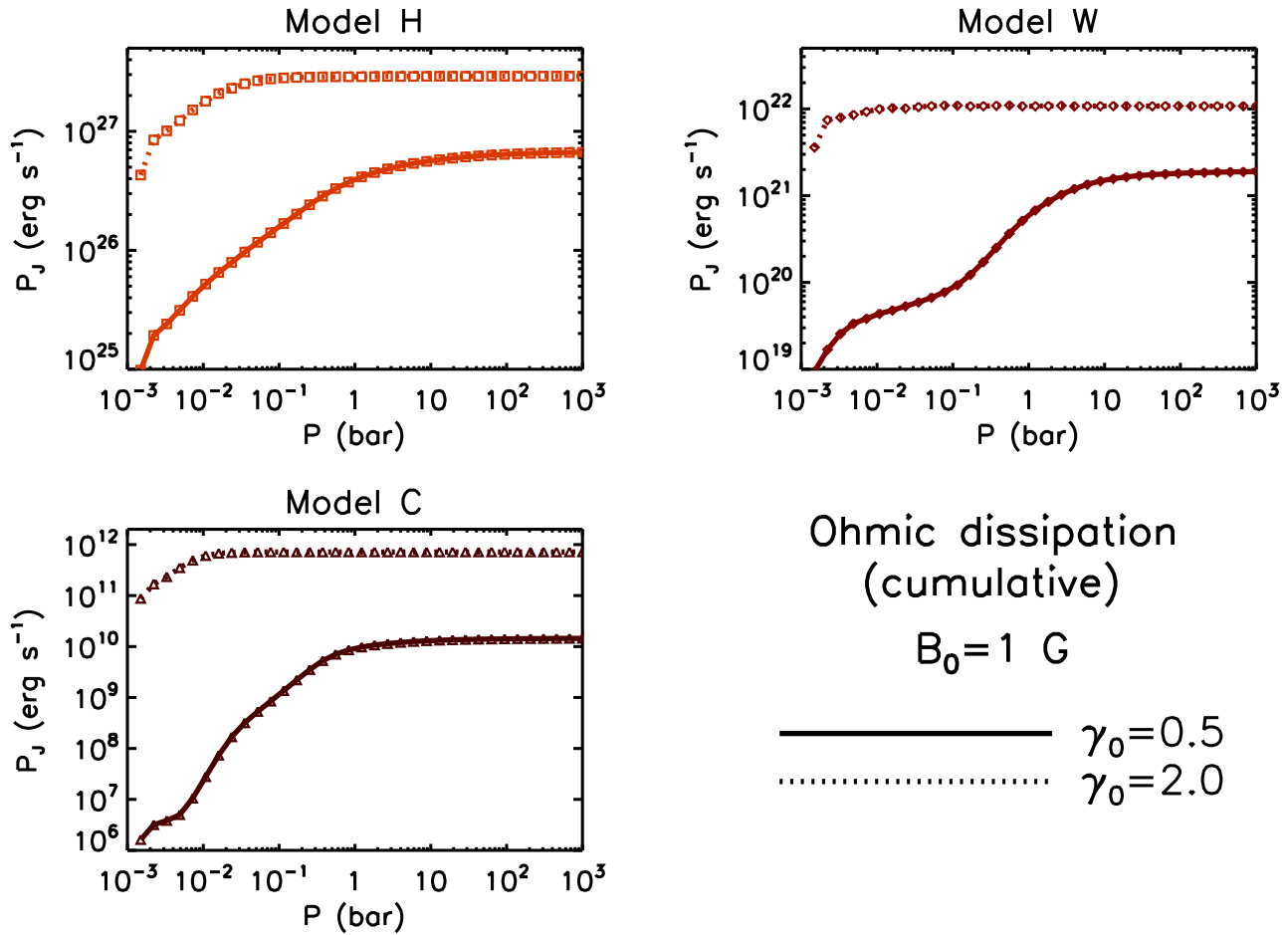}
\caption{Dissipated Ohmic power as a function of depth, for the three representative
models C, W, H, both with and without temperature inversions. A 
magnetic field strength of 2~G at the poles has been assumed. Note that the penetration
depth increases with the strength of the irradiating flux; for the same $T_{\rm irr}$,
it is larger in models without temperature inversions.}
\label{fig:ohmic}
\end{figure*}

\subsubsection{Dissipation from shocked gas}

In order to estimate the presence and amount of dissipation due to shocked
gas, we begin by keeping track of the Mach number at each location within the
flow; since zonal velocites are the most significant ones, we define the Mach number as
\begin{equation}
{\cal M}(\Theta,\Phi,r)\;=\; \frac{v_\Theta}{c_s}\,.
\label{eq:mach}
\end{equation}
Shocks are likely to occur for ${\cal M}\ga 1$, barring special conditions which allow the pre- and post-shock regions to remain in causal contact \citep{zr66}.  One may regard our simple estimates of the shock dissipation as upper limits since ${\cal M}\ga 1$ is a necessary but insufficient condition for the development of shocks.

At each pressure level within the atmosphere, flow velocities change with
both latitude and longitude. To investigate how deep shocks can penetrate, in
Fig.~\ref{fig:mach} we show the maximum Mach number as a function of 
atmospheric depth, for the three representative models C, W, and H, each with and
without temperature inversions. In the coldest model, the flow is always subsonic.
Wind speeds increase, as expected, with the strength of the irradiating flux and hence with 
the day-night temperature gradients. In the W model, the flow
can become supersonic (with Mach numbers of $\sim 1.5-2$)
at pressure levels above a few tens of bars. In the H model, shock penetration
reaches a few bars, with Mach numbers $\sim 2.5-3$. For the same strength of irradiation,
flows with temperature inversions reach a higher Mach number, due to the higher 
temperatures attained in the outermost layers.

In order to make a crude estimate of the amount of energy available to
be dissipated in shocks, we first compute the total amount of kinetic
energy available in the flow up to pressure $P$, $E_{\rm
  kin}(P) =(1/2)\int_{r(P_0)}^{r(P)} dV \rho (v^2_\Theta + v^2_\Phi)$.
Note that the dominant term comes from the zonal component ($v_\Theta$).
Figure \ref{fig:energy} shows $E_{\rm kin}(P)$
in the three representative models C, W and H. The cumulative amount
from the whole flow varies by over three orders of magnitude between
the C and the H model, with a weaker dependence on the shortwave opacity.  
Generally, we find that the stronger the irradiation, the deeper the penetration 
of the zonal winds.

Together with the local kinetic energy density, we track, at each
point of the flow, both the Mach number and the Richardson number.  If
either ${\cal M}>1$ or ${\cal R}i<1/4$, we assume that a fraction of
0.5 of the kinetic energy is dissipated/converted into heat. In strong shocks, this
fraction is given by $4(\gamma-1)/(\gamma+1)^2$, where
$\gamma$ is the adiabatic index \citep{dm93}; for $\gamma=7/3$, this amounts to
0.48.  For shear instabilities, Li \& Goodman (2009) estimate the
dissipated fraction of kinetic energy to be $\approx 0.5$--1.  Hence we
take $E_{\rm dis}(P) = (1/2) f_{\rm dis}\int_{r(P_0)}^{r(P)} dV \rho
(v^2_\Theta + v^2_\Phi)$, where $f_{\rm dis}=0.5$ if ${\cal
  M}>1$ or ${\cal R}i<1/4$ and $f_{\rm dis}=0$ otherwise.  We remark
once again that this is a rather crude estimate, motivated by the fact
that current global 3D circulation models are not yet able to follow
the development, growth, and dissipation of instabilities and shocks.

The cumulative dissipated energy, $E_{\rm dis}(P)$, is
shown in Fig.~\ref{fig:energy} for the C, W and H models.  In the C model, $E_{\rm
  dis}(P)=0$ since, as discussed above, the criterion for the onset
of the KH instability is never satisfied, and the velocities are always
subsonic. In the W and H models, on the other hand, a fraction of
about 10-20\% of the kinetic energy is available for dissipation, with
the fraction being generally larger in models with temperature
inversions, due to the stronger velocities (and their gradients) in the
upper atmospheric layers, which are the ones more prone to becoming
either KH unstable or shocked. For a given value of $\gamma_0$, 
dissipation occurs down to deeper levels as the strength of the
irradiating flux increases, reflecting the penetration depth of the
shocks (cf. Fig.~\ref{fig:mach}). However, even in the hottest models,
shock dissipation never penetrates deeper than a few bars.

\subsection{Ohmic dissipation}

Previous work (Batygin et al. 2010; Perna et al. 2010b) has shown
that, for Jupiter-like magnetic field strengths, Ohmic dissipation can
be a significant source of energy for the exoplanet. In particular, using
a simulation specialized to the physical parameters and conditions of
HD 209458b, Perna et al. (2010b) showed that Ohmic dissipation can
penetrate down to levels deep enough to influence the radius evolution
of the exoplanet. Since irradiation is the driver of the flow dynamics,
currents (and hence Ohmic dissipation) are expected to depend
sensitively on its strength. In the following, we investigate this
dependence using our 16 baseline models.  We assume the $B$-field to
be an axisymmetric dipole where the rotational and dipolar axes are aligned.  The dependence of
Ohmic dissipation on the magnitude of the $B$-field has been explored
(inclusive of an approximate feedback from magnetic drag), by Perna et al. (2010b).

To compute Ohmic dissipation within the context of our simulations, we follow
the formalism of Liu et al. (2008) and Perna et al. (2010b). Under the assumption that zonal winds
are stronger than the meridional motions, the only relevant component of the
induction equation is the toroidal one.
Under steady-state conditions, it was showed by Liu et al. (2008) that the 
resulting dominant component of the current induced by the zonal flow is the meridional one,
which we write below in terms of the latitude $\Phi$
and the longitude $\Theta$, and also
incorporating the expression for an aligned dipole for the magnetic field:
\begin{eqnarray}
{\cal J}_\Phi &=& \frac{B_0 r_{\rm max}^3 \sigma_e}{\left( r_{\rm max}
  - \tilde{z} \right) c} \int^{\tilde{z}}_0\hspace{-0.05in}dz^\prime \left[\frac{2
  \sin\Theta}{\left( r_{\rm max} - z^\prime \right)^2} \frac{\partial
  v_\Theta}{\partial z^\prime} + \frac{2 v_\Theta \sin\Phi}{\left(
  r_{\rm max} - z^\prime \right)^3}\right. \nonumber \\ 
&+& \left.\frac{\cos\Phi}{\left(
  r_{\rm max} - z^\prime \right)^3} \frac{\partial v_\Theta}{\partial
  \Phi} + \frac{v_\Theta \sin\Phi}{\left( r_{\rm max} - z^\prime
  \right)^3}\right].
\label{eq:j}
\end{eqnarray}
Here, $\eta=c^2/(4\pi\sigma_e)$ is the electrical resistivity, and the
electrical conductivity $\sigma_e$ is computed using the same
formalism as in Perna et al. (2010a,b).  The familiar radial
coordinate, denoted by $r$, is equal to $z+R_p$ where $z$ is the
vertical height measured from the bottom of the simulation domain.
The top of the simulation domain is located at $r_{\rm max} =
\mbox{max}(z+R_p)$.  We choose to compute the current density as a
function of the variable $\tilde{z}$, which is defined such that
$\tilde{z}=0$ when $r = r_{\rm max}$, i.e., it is the vertical height
measured from the top of the model atmosphere.  
Again, $v_\Theta$ is the zonal velocity.  The quantity $B_0$ is the
normalization term in the dipole magnetic field, such that it is $B_0$
at the equator and $2B_0$ at the poles. For our calculations, we choose
$B_0=1$~G. Note that in the above
equation we have neglected the contribution from an unknown boundary
current.

The total Ohmic power dissipated between the bottom shell at pressure $P_0$ and a shell
of material located at pressure $P$, is then readily computed as

\begin{equation}
P_J(P)=\int_{r(P)}^{r(P_0)}\hspace{-0.05in}dr' {r'}^2 \int_0^{2\pi}\hspace{-0.05in} 
d\Theta\int_{-\frac{\pi}{2}}^{\frac{\pi}{2}}\hspace{-0.05in} d\Phi
\cos\Phi \frac{[{{\cal J}_\Phi(r',\Theta,\Phi)}]^2}{\sigma_e(r',\Theta,\Phi)}.
\label{eq:PJ}
\end{equation}

Fig.~\ref{fig:ohmic} shows the cumulative Ohmic power as a function of
the atmospheric depth, for the three representative models C, W and H,
both with and without temperature inversions. The magnitude of $P_J$
has a very strong dependence on the irradiating flux, varying at
maximum by about 16--17 orders of magnitude between the C and the H
models.  While the increasing zonal wind speeds with $T_{\rm irr}$ do contribute to 
this effect, the dominant factor is the exponential dependence of the electrical 
conductivity on temperature.

For the same $T_{\rm irr}$, the maximum Ohmic power is larger for
$\gamma_0=2$ (with a temperature inversion) than it is for $\gamma_0=0.5$ (no
inversion), while the penetration depth on the other hand is larger
for $\gamma_0=0.5$. This behaviour can be understood as follows. For
$\gamma_0=2$, the outer layers of the atmosphere are hotter than the
inner layers, and hence the conductivity is much larger in the outer
atmospheric regions.  Since velocity gradients are also stronger in
these regions, the net result is the generation of very large
peripheral currents and hence large dissipation, which however
saturates, even in the hottest models, close to the photosphere. For
$\gamma=0.5$ on the other hand, the outer layers are colder, and the
low conductivity results in weak currents. Deeper in, the temperature
rises, and so does the conductivity, resulting in an increase of the
Ohmic power at deeper atmospheric levels.   However, the maximum dissipated 
power saturates at smaller values than than for the case of
$\gamma_0=2$, since for $\gamma_0=0.5$ the region with higher
conductivity corresponds to zonal flows with weaker gradients.

\begin{figure}
\centering
\includegraphics[width=9cm]{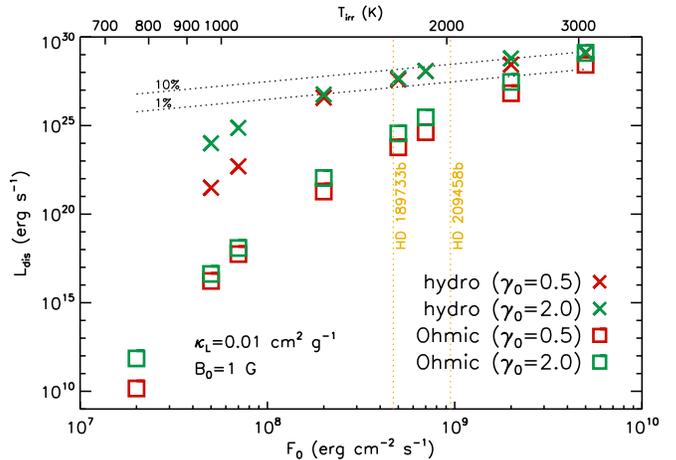}
\caption{Comparison between hydrodynamic and Ohmic dissipation, for
  all of the models presented in our study. For the hydrodynamic dissipation, we assume
  that the energy is dissipated over a rotational timescale.  The
  Ohmic power is computed for a dipolar $B$-field with strength of 2~G
  at the pole. Without feedback from magnetic drag, the Ohmic power
scales as $B_0^2$.  The dotted lines labelled ``1\%" and ``10\%" correspond to 1\% and 10\% 
of the stellar irradiation, respectively.}
\label{fig:power}
\end{figure}

A comparison between the total power (i.e., integrated over the entire
atmosphere) generated through shocks and shear instabilities, and
through Ohmic dissipation, is shown in Fig. \ref{fig:power}.  Since
our simulations do not explicitly follow shocks and the development of
shear instabilities, they cannot make predictions for the timescales
over which dissipation associated with these phenomena occur. To
compute the hydrodynamic power in Fig.~\ref{fig:power}, we assumed
that $E_{\rm dis}$ is dissipated over a rotational timescale (which is
comparable in magnitude to the advection timescale). The 2D
simulations by Li \& Goodman (2010) show that, once shocks are formed,
they slow down over a very short time (shorter than the advection
time).  As long as the dissipation timescale is shorter than
$2\pi/\Omega_p$, the power shown in Fig.~\ref{fig:power} can be
considered as a good proxy for its corresponding time-averaged
quantity over a rotational cycle\footnote{Lacking 3D evolutionary
  planet models which can assess the effect of anisotropic heat
  injection in the planet interior on the thermal evolution of the
  planet, for an order-of-magnitude comparison with the 1D
  evolutionary models of Guillot \& Showman (2002), we consider the
  averaged dissipated hydrodynamic power over a rotational cycle.}.

Our simulations show that the Ohmic power is a much steeper function
of irradiation than the hydrodynamic power, which we find to be
dominated by shocks.  For the latter, only velocities matter, while
the dissipated Ohmic power depends on velocity gradients. But most
importantly, as the temperature increases, so does the ionization
fraction, and hence the conductivity, which results in stronger
currents. However, we remark, once again, that the slope at high $T_{\rm irr}$
would be shallower if magnetic drag were included self-consistently in the
simulations.  Specifically, \cite{menou12b} has shown, via simple scaling laws, 
that magnetic drag rapidly retards the atmospheric winds for strong stellar irradiation, 
and that, for a given magnetic field strength, there is a value of the irradiation/equilibrium 
temperature for which Ohmic dissipation peaks (and then declines as irradiation becomes 
more intense).  Our results for the dissipated power should therefore be considered as 
upper limits. 

Finally, we note that \cite{heng12} provides a companion study to our paper by using semi-analytical models to elucidate the interplay between atmospheric scattering, absorption and Ohmic dissipation with relevance to the possible presence of atmospheric clouds or hazes.

\subsection{Implications for the problem of radius inflation}

As discussed in \S1, the anomalously large radii of a fraction of hot
Jupiters remains a persistent problem for researchers.  An ``extra"
source of internal energy is needed for the radii to be larger than
predicted by standard evolutionary models.  Guillot \& Showman (2002)
argue that, if about 1\% of the stellar insolation flux were deposited
at pressures of tens of bars deep in the atmosphere, cooling would
slow down sufficiently to explain the inflated radii of hot Jupiters.
The evolutionary model presented by Guillot \& Showman (2002) was
tailored to the specific case of the exoplanet HD~209458b, with a
radius of $R=1.35R_J$ ($R_J$ is the Jupiter radius), and a
convective-radiative boundary located at pressures of about 1 kbar (at
its current age).  Note that, for exoplanets with comparable mass, the
boundary layer between the convective/radiative zones is located at
lower pressures as the planet is younger and larger (e.g., Guillot \&
Showman 2002, Burrows et al. 2003). The amount of internal heat needed to inflate the
  radius by a certain $\Delta R$ is also a strong function of the
  planet mass (Miller, Fortney \& Jackson 2009). The smaller the
  mass, the easier it is to inflate the planet. For example, Miller et
  al. (2009) show that a 20\% increase in radius (with respect to
  $R_{\rm J}$) can be achieved with a mere $10^{24}$~erg~s$^{-1}$ in a
  planet of mass 0.1~$M_{\rm J}$, while the same radius increment would
  require about $10^{28}$~erg~s$^{-1}$ for a 10~$M_{\rm J}$ planet.

While several mechanisms could transport and dissipate energy into the
interior, such as eccentricity damping (Bodenheimer et al. 2001), or
forced turbulent mixing in the radiative layer (Youdin \& Mitchell
2010), here we have investigated energy deposition resulting from the
development of shear instabilities and shocks, and from Ohmic
dissipation. For all of these phenomena, our numerical simulations,
while not modeling the dissipation itself and its feedback on the
flow, do allow us to estimate their importance from an energetic point
of view, as a function of the strength of the irradiating flux and of
the atmospheric opacity.

We have found that shear instabilities can develop only very
superficially, even in the hottest models, and hence they are unlikely
to meaningfully influence the energy budget of the exoplanet.  
If the source of weak turbulence postulated by \cite{ym10} results from shear
instabilities, then their proposed mechanism of forced turbulent
mixing does not operate deeply enough.  To further test the relevance
of their mechanism in global circulation models requires the
specification of an instability and its criterion for being invoked.

Shocks penetrate deeper, up to a few bars in the hottest models. This is a
depth which could be interesting for influencing the thermal evolution
of planets, and hence the degree by which their radius is able to
shrink during the evolution. For the particular case of HD~209458b,
the evolutionary models of Guillot \& Showman (2002) showed that, even
at a pressure as shallow as 5 bars, dissipation of 10\% of the
absorbed stellar flux (corresponding to $\dot{E}=2.4\times
10^{28}$~erg~s$^{-1}$ for this planet) is sufficient to retard cooling
so that the planet contracts to its current size on a timescale
$\sim$~5~Gyr, comparable to its age. Inspection of the trend for the
dissipated hydrodynamic power versus irradiating flux in
Fig.~\ref{fig:power}, shows that, for the strength of ${\cal F}_0$
corresponding to the case of HD~209458b, the hydrodynamic power is on the
order of a few $\times 10^{28}$~erg~s$^{-1}$. Therefore, since the
dissipated hydrodynamic power that our models predict falls in an
interesting order of magnitude range (at least for the most irradiated
objects), and its penetration depth also reaches levels close to the
interesting ones identified by evolutionary models\footnote{It is also
  important to note that vertical advection can help transport the
  dissipated energy to deeper levels. In our models, the vertical
  advection timescale is $\sim 10^5$ s, comparable to the
  advection timescale; for an HD~209458b-like planet, \cite{hfp11} showed that vertical mixing extends down to about 10~bars.},
we suggest that shock dissipation could indeed play a significant role
in the thermal evolution of the planets, and hence influence their
radii, although especially so in the most irradiated planets.

Our investigation of Ohmic dissipation has showed that, for a
$B$-field strength on the order of a Gauss, the total dissipated Ohmic power is
generally lower than the hydrodynamic power except for the most
irradiated hot Jupiters. However, it penetrates deeper, down to
pressures of several tens of bars, if the atmosphere has no
temperature inversion.  In the evolutionary scenario of Guillot \&
Showman (2002), a deeper penetration depth can be ``traded" for a
smaller power. For example, in the case of HD~209458b, their
evolutionary models could reproduce the exoplanetary radius at its current
age both with 10\% of the irradiating flux dissipated at 5~bars, or
with 1\% (i.e., $2.4\times 10^{27}$~erg~s$^{-1}$) dissipated at
21~bars.  For a $B$-field strength of a few Gauss, this power is available in the
most irradiated objects. Clearly, the importance of Ohmic dissipation
for the thermal evolution of planets also depends strongly on the magnitude
of the $B$-field, since the Ohmic power scales as\footnote{As discussed in \S 4.2, this scaling
would be shallower if feedback effects from magnetic drag
were accounted for, but it would still remain an increasing function
of $B_0$ for a wide range of irradiating fluxes.} $B_0^2$, and, for
fixed planet conditions, the penetration depth also increases with
larger $B_0$ (Perna et al. 2010b).  However, for a given $B$-field,
since both the penetration depth and the power dissipated at a fixed
pressure are rapidly increasing functions of the strength of the
irradiating flux, the influence of Ohmic dissipation on radius
inflation is expected to depend strongly on stellar irradiation.

\section{Qualitative Comparison with Observations}

Observations of hot Jupiters have allowed to characterize, albeit with
varying statistical level, the atmospheric properties that we have
discussed so far, and in particular, the efficiency of heat
redistribution, the hotspot offset and the exoplanetary radius.  Here we do not
attempt to make any detailed, quantitative comparison between our models and the
data; the former have been constructed for ``typical'' values of the exoplanetary
parameters, and hence they are not generally applicable to a specific object.  
For the same value of the irradiating flux, a scatter is
expected due to the different intrinsic properties of the exoplanet, of
which here we have only examined the dependence on the shortwave
opacity for two values.  (See \citealt{heng12} for a broader exploration of $\gamma_0$ values, and also the influence of shortwave scattering, using semi-analytical models.)  The data on the other hand, still suffer from
measurement uncertainties which would make quantitative comparisons
with models not fully reliable yet.  However, the {\em qualitative
  trends} in both the data and the model are fairly robust, and these
are the ones that we are interested in reproducing at this stage.

\subsection{Data summary}
Figure \ref{fig:obs1} summarizes the main observational constraints on the radius
inflation and atmospheric circulation of hot gas giant exoplanets. The
exoplanetary radius measured for transiting gas giants ($M>0.3 M_J$) is
shown as a function of the irradiation temperature. Different symbols indicate more/less massive exoplanets,
exoplanets with or without a temperature inversion measured near the
infrared photosphere, and exoplanets with or without efficient day-night
temperature redistribution. The presence of a temperature inversion is
inferred from the broadband infrared emission spectra measured during
secondary eclipse with the {\em Spitzer} satellite. The efficiency of
the day-night temperature redistribution is inferred from the
comparison of the effective temperature in the {\em Spitzer} passbands with
the equilibrium temperature. If the measured temperature is much
larger than the equilibrium temperature, the redistribution of heat to
the night side is likely to be low\footnote{This inference is tentative,
since the albedo and the redistribution of flux in frequency by
molecular spectral features will also play a role. The indications for
specific objects may turn out to be incorrect, but hopefully this will
simply introduce some random noise and not entirely mask the existing
correlations between the efficiency of redistribution and other
parameters.}. In one case, HD 189733b, the heat redistribution is
measured more directly, from the infrared phase curve of the exoplanet at 8
and 24 microns (Knutson et al. 2007, 2009).

The exoplanet radii and irradiation data have been taken from the
edited compilation at \texttt{www.inscience.ch/transits} (see
references therein for primary sources).  Information on the
absence/presence of temperature inversions is adopted from Knutson et
al.  (2011, see references therein), while information on
redistribution is taken from the compilation of Harrington (2011),
updated with more recent data from {\em Spitzer} for HD80606, XO-3,
HAT-P-7 and WASP-12; we also include inferences on XO-4, HAT-P-6 and
HAT-P-8 (Todorov et al. 2011). It is important to note that 
inferences on redistribution (presented in the lower
panel of Figure 8) are based on data that stretches the capacity of the
{\em Spitzer} instruments to their limits, and that the uncertainties
associated with instrumental effects are still high. In at least two
cases, Ups~And~b and HD~149026b, further measurements have shown the
initial estimates for the surface temperature and redistribution
efficiency to be completely incorrect. These two objects are not in
our plot, Ups~And~b because it is not a transiting planet and
therefore its radius is not measured, and HD~149026b because it is
not in synchronous rotation and is a core-dominated planet rather
than a hot Jupiter, and therefore it does not fit our model assumptions.

Information on the angular offset of the hotspot is only sparsely available.  
{\em Spitzer} observations of HD~189733b show
that the hottest region of the thermal phase curve is
advected to the east of the substellar point (Knutson et al 2007) by
$16 \pm 6$ degrees; Knutson et al. (2009) inferred the 1D brightness maps 
at 8 and 24 microns and found the offsets to be $30 \pm 4$ and $23 \pm 7$ degrees, respectively.  
Measurements for a couple of other objects are
less certain.  Crossfield at al. (2010) report, for Ups And b,
which has a good redistribution, an offset of the 1D brightness map to the east of either
$57\pm 21$ degrees or $84.5\pm 2.3$ degrees, using two different data
sets for the analysis.  Cowan et al. (2012), report, for WASP-12b, an offset 
to the east of $12\pm 6$ degrees from the {\em Spitzer} phase curve at 
4.5 microns, and an offset of $53\pm 7$~degrees at 3.6 microns.
 
\subsection{Observational vs. theoretical trends}

\begin{figure}
\centering
\includegraphics[width=8.5cm]{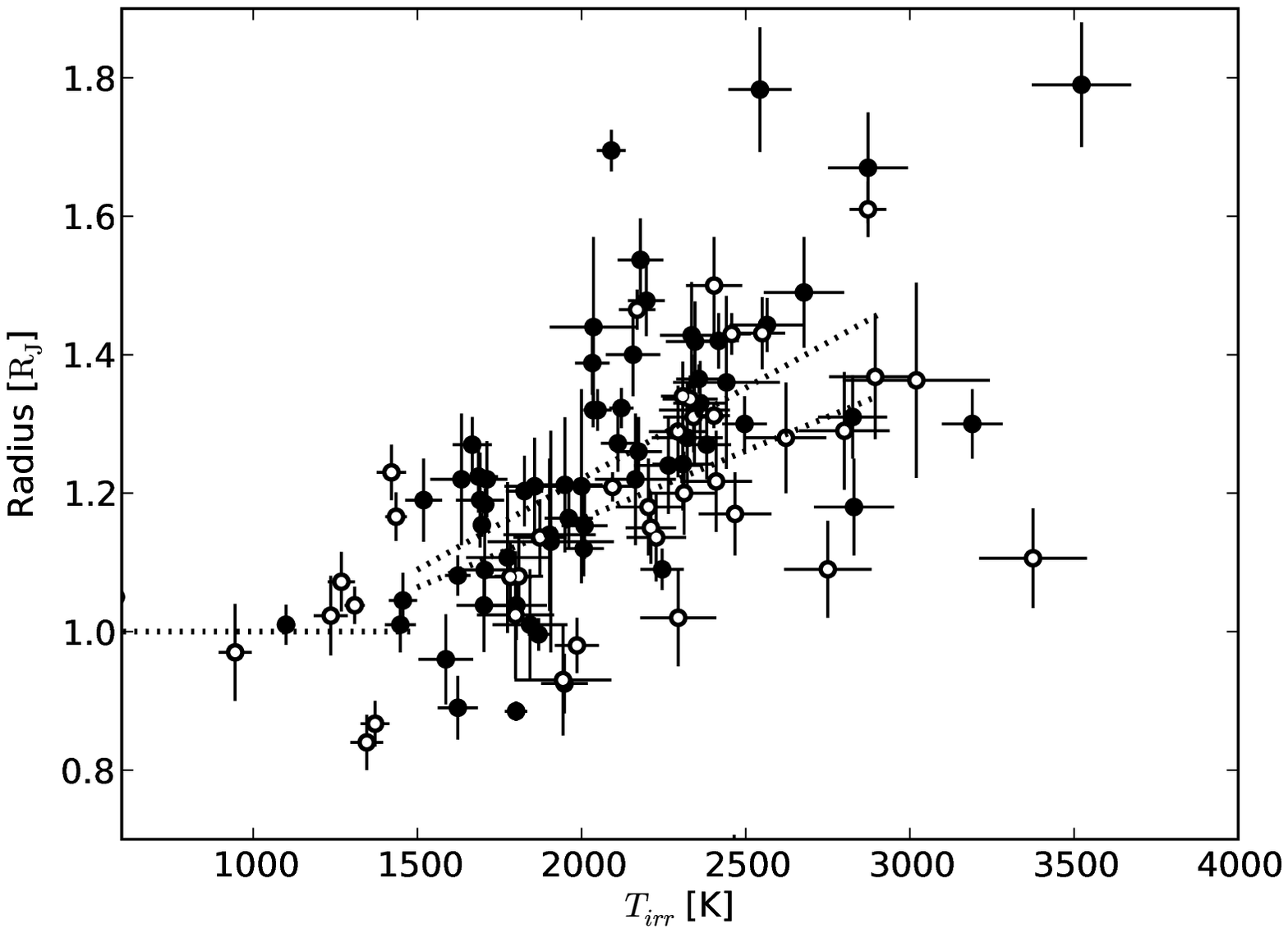}\\
\includegraphics[width=8.5cm]{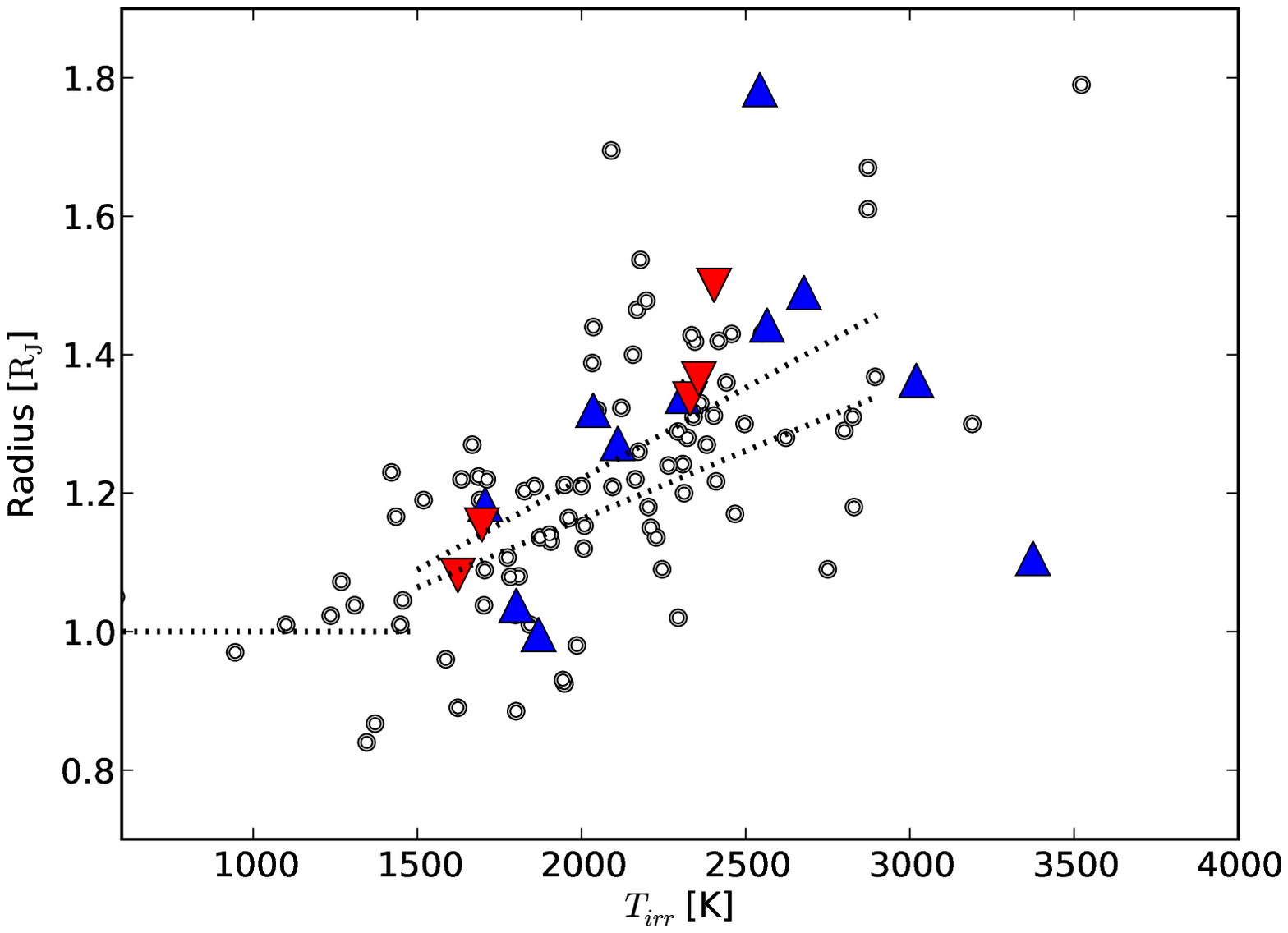}\\
\includegraphics[width=8.5cm]{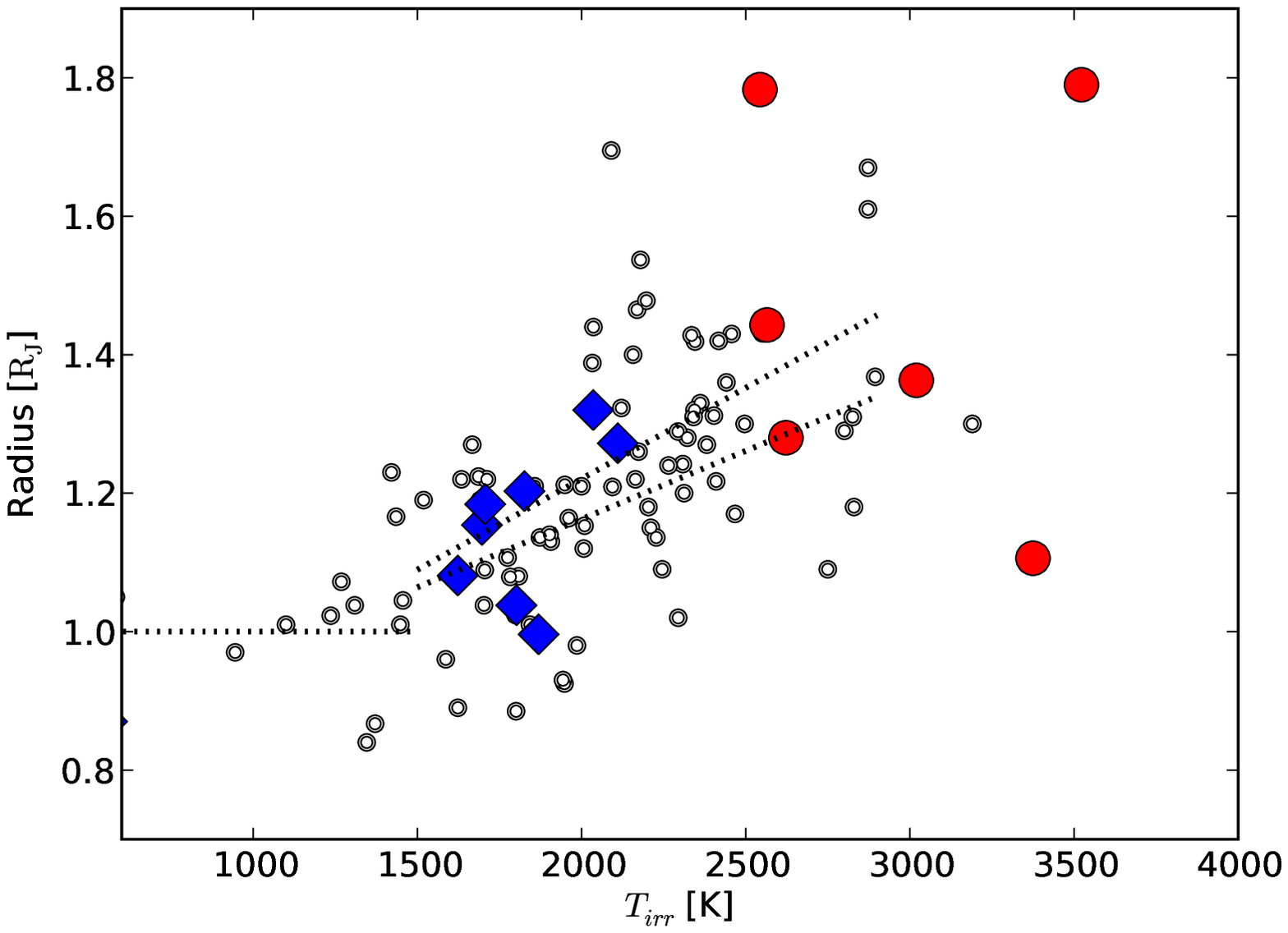}
\vspace{0.1in}
\caption{{\em Upper panel:} Exoplanetary radius as a function of
  irradiation temperature for transiting gas giants. Exoplanets in the
  range $0.3<M<1.5 M_J$ are shown as closed symbols, exoplanets
  heavier than 1.5 $M_J$ as open symbols (data from
  \texttt{www.inscience.ch/transits}).  Dotted lines show
  the temperature range where no significant radius anomaly is observed
  (flat line), and least-square linear fits to the low-mass data and
  high-mass data separately.  {\em Middle panel:} Same data as in the
  upper panel, but with upside-down red triangles indicating
  inverted temperature profiles, and blue triangles normal
  profiles. {\em Bottom panel:} Same data as in the other two panels,
  but with blue diamonds indicating probable efficient redistribution,
  and red circles probable strong day-night contrast.  }
\label{fig:obs1}
\end{figure}

Figure \ref{fig:obs1} illustrates the strong correlation between irradiation and
radius inflation. This indication that injection of a fraction of the
irradiation into the exoplanet as internal entropy is responsible for the
radius inflation (as first suggested by Showman \& Guillot 2002) is
reinforced by the observed correlation between radius inflation and
the mass of the exoplanet (Fortney et al. 2011), and by the absence of
radius inflation for cooler Jupiter-sized exoplanet candidates from the
Kepler mission (Demory \& Seager 2011). Hence, the observations
strongly support explanations of radius inflation as a consequence of
high stellar irradiation. The observed slope of the dependence,
though, seems even steeper than that predicted by the injection of a
constant fraction of the stellar flux as internal entropy as proposed
by Showman \& Guillot (2002), suggesting that the efficiency of the
coupling mechanism increases with temperature.
There is no significant correlation between radius inflation and
orbital eccentricity (Pont 2009), nor with temperature inversion.

For the dissipation mechanisms that we have explored in this work,
both of hydrodynamic and of magnetohydrodynamic origin, our
simulations show a strong correlation with the strength of the
irradiating flux. In particular, for a $B$-field on the order of a Gauss, we found
that, for $T_{\rm irr}\la 2000$~K, the dissipated Ohmic power is
too small (and its penetration depth too shallow), to affect the
radius evolution in any appreciable way. However, as $T_{\rm irr}$
approaches $\sim 2000$~K, the Ohmic power grows noticeably, and its
penetration depth, reaching down to several tens of bars, can then
affect the evolution of the exoplanetary radius (Guillot \& Showman
2002).  Interestingly, as Fig.~\ref{fig:obs1} shows, radius inflation
starts to become important as the irradiation temperature approaches
the 2000~K range.  We also note that variations of the magnetic field
from exoplanet to exoplanet can result in a substantial scatter in the
radius evolution, since the Ohmic power scales as $B^2_0$, and this
rapid growth with magnetic field strength dominates over the reduction
to the power caused by the feedback effect of magnetic drag, at least 
for an interesting range of $B$-field strengths (Perna et
al. 2010b). For a Jupiter-like magnetic field ($\sim 14$~G at the pole),
the power produced by Ohmic dissipation can already affect planets with
$T_{\rm irr}\sim 1700-1800$~K. 
The atmospheric (shortwave) opacity further contributes to the
scatter, since the penetration depth of Ohmic dissipation is a non-negligible
function of this variable.

In summary, and at a qualitative level, our simulations predict that
both hydrodynamic dissipation (dominated by shocks) and Ohmic
dissipation (for magnetic field strengths at least as large as
several Gauss), can play an important role in altering the thermal
evolution of hot Jupiters, slowing down cooling, and favoring the
presence of planets of larger dimensions than predicted by
evolutionary models without ``extra" energy injection.  A significant
dissipated power (both hydrodynamic and Ohmic if $B_0\ga 7-8$~G), on the order of
several percents of the stellar insolation, is predicted for irradiation
temperatures $T_{\rm irr}\ga 1700-1800$~K.  For both forms of
dissipation, the stronger the irradiating flux, the larger the total
dissipated power and the deeper the penetration depth, resulting in an
increasingly larger inflation as the flux becomes more intense.
At large insolation, even if the direct dissipation depth is rather shallow
(as it is for shocks and for currents in highly inverted temperature profiles),
vertical mixing can help to carry down heat to levels of tens of bars, hence
maintaining the energy injection in the relevant regions for affecting
the thermal evolution of the planet. 

A quantitative prediction of the slope of the $R_p(T_{\rm irr})$ curve
would need a coupling between a specific dissipation mechanism and an
evolutionary model for the exoplanet, which is beyond the scope of the 
present study.  But even so, as discussed above,
a large scatter would be expected due to exoplanets with different
intrinsic properties (opacity, gravity, magnetic field strength, etc).

Although the statistical information is still rather limited,
inspection of Fig.~\ref{fig:obs1} further shows that the efficiency of
day-night heat redistribution seems to diminish rapidly for $T_{\rm irr}$
above 2200 K, with the hottest gas giants showing day-side
temperatures in the infrared much higher than the equilibrium
temperature, an effect pointed out by Harrington (2011).
Interestingly, our simulations (cf. Fig.~\ref{fig:contrast}) also
show that, from irradiation temperatures around 2200--2500 K, the atmospheric
flow pattern begins to evolve from a very good redistribution
to a regime of no-redistribution.  Exoplanets with temperature
inversions evolve more quickly towards a no-redistribution
regime by displaying, for the same irradiating flux, a larger
day-night flux contrast. This is again supported by observations,
as well as by other work (Fortney et al. 2008). 
It may also be worthwhile to note that the drop in redistribution
efficiency appears to occur around the temperature at which radius
inflation saturates, although this is only a tentative indication at
this point due to the low number of known cases, the large scatter at
high temperatures, and the assumptions involved in inferring
redistribution from infrared day-side fluxes in the {\em Spitzer}
passbands. From a theoretical point of view, a saturation of radius
growth at very high values of $T_{\rm irr}$ is expected due to
feedback effects (e.g., shock dissipation and magnetic drag, both
contributing to reduce the speeds of the fastest flows).

For the hotspot offset with respect to the substellar point, there is
insufficient data statistics yet to compare theoretical and
observational trends.  However, we do note that, as the observations
show that the offset can be both rather small or very large, our
theoretical modeling also shows a wide range of values, and further
predicts a correlation between offset width and degree of
redistribution. Although our modeling here has been for ``typical"
exoplanets and is not tailored to the properties of specific objects,
it is however tempting to point out that, with the irradiating flux
level of HD~189733b, our predicted offset is in the
$20^\circ$--$40^\circ$ range (cf. Fig.~\ref{fig:offset}), consistent
with the range of values ($16^\circ$--$34^\circ$) obtained from the
inferred 1D brightness maps, at 8 and 24 microns, by Knutson et
al. (2009).  In the case of Wasp-12b, the extremely high value of
  the irradiation temperature, $T_{\rm irr}\sim 3500$~K, is above the
  temperature range studied in our simulations. However, inspection of the offset
  trend at high $T_{\rm irr}$ in Fig.~\ref{fig:offset} would suggest a
  value around a couple of tens of degrees at those high temperatures.  The measured offset of
  $16\pm 4$ degrees at 4.5 microns appears compatible with this
  estimate.  On the other hand, the much larger offset of $53\pm
  7$~degrees at 3.6 microns does not.  The difficulty in reconciling this
  large phase offset with the other properties of the object was
  already discussed by Cowan et al. (2012), though they also noted
  that the measurement could be affected by highly-correlated
  residuals near the purported peak. If real, such a measurement might
  indicate a rather unusual opacity, and a detailed theoretical
  modeling would require a non-grey radiative
  transfer treatment (Fortney et al. 2006, Showman et al. 2009, Burrows et al. 2010). Similarly,
for the case of Ups And b, the unusually large offset
values reported by Crossfield et al. (2010), albeit with somewhat
large error bars, cannot be simply accounted for with the
opacity values adopted here. However we note that, even within the
simple framework of the dual-band model, larger offsets can be
produced for smaller values of both the optical and infrared opacities.

\section{Summary}

Irradiation by the parent star is the main driver of the radiative and
dynamical properties of hot Jupiters.  Opacity in the exoplanetary
atmosphere further plays an important role in determining the response
of the atmospheric flow to a certain amount of irradiating flux. 

In this paper, by means of 3D atmospheric circulation models with dual-band radiative
transfer, we have explored the role of
irradiation and opacity in determining some of the main observational
properties of hot gaseous exoplanets, with particular emphasis on heat redistribution
and on hydrodynamic and magnetohydrodynamic dissipation. 
Although our simulations are still idealized in several respects
(simplified radiative transfer, constant opacities, lack of feedback
effects due to dissipation of instabilities and shocks and to
magnetic drag), they still allow us to capture some broad
trends exhibited by the observations.

Our results are summarized below:
\begin{itemize}
\item
For irradiation temperatures $T_{\rm irr}\la 2200-2400$~K, we find
heat redistribution to be very efficient.  Near-perfect redistribution
(i.e., $F_{\rm day}/F_{\rm night}\approx 1$) only occurs up to $T_{\rm irr}\la 1500$~K,
but in the range  1500~K~$\la T_{\rm irr}\la 2400$~K the day-night flux contrast 
is still relatively small ($\lesssim 2$). As the stellar irradiation becomes more intense, 
redistribution begins to fail. 
Observations, albeit limited by the low statistics, hint at the redistribution
breaking down around $T_{\rm irr}\ga 2200-2400$~K, in broad agreement
with our theoretical results.
Our simulations support the simple intuitive result that it is
the interplay between advection and radiative cooling which determines the effectiveness of heat redistribution.
Redistribution begins to falter as the advective timescale becomes
much longer than the radiative timescale.

\item
For the same strength of irradiation, exoplanets with a temperature
inversion (here parameterized by means of a larger shortwave opacity)
display a higher day-night contrast than exoplanets with no
temperature inversion, since starlight is deposited higher up in the 
layers of the atmosphere dominated by radiative cooling (rather than advection).

\item
The offset of the hottest region (hotspot) from the substellar point
can exhibit a wide range of values; it is generally very large (even
entering the night side) for exoplanets with very efficient redistribution
(low $T_{\rm irr}$) and it moves closer to the substellar point with
increasing strength of the irradiating flux. For a fixed value of the 
irradiating flux, opacity further plays a role in determining the offset. 
The smaller the shortwave opacity, the larger the offset for the same $T_{\rm irr}$.

\item
A measurement of the Richardson number throughout the 3D flow showed
that the Kelvin-Helmholtz instability is unlikely to play a major role
in the thermal evolution of the planet. We found that the the
criterion for the instability onset, ${\cal R}_i<1/4$, is satisfied
only in the uppermost layers of the flow, down to several millibars
(if at all).

\item
Shocks, associated with the presence of supersonic flows
(up to Mach numbers of $\sim 2.5-3$ in the hottest models), can dissipate
up to 10-20\% of the available kinetic energy. The total dissipated
power over an advective timescale reaches several percent of the insolation
energy power for irradiation temperatures $T_{\rm irr} \ga 1700-1800$~K, and
it then increases with $T_{\rm irr}$. Hence, for the most irradiated objects,
the magnitude of the hydrodynamic power is such that it could noticeably influence
the thermal evolution of planets. 
However, the direct penetration depth of the dissipation does not extend
 beyond a few bars even in the most irradiated hot Jupiters. 
Some amount of vertical advection would then be necessary to bring the heat to deeper levels.
In the most irradiated objects, our simulations predict vertical mixing to 
penetrate down to tens of bars (see Appendix \ref{append}), which could make hydrodynamic dissipation
an interesting contributor to the thermal evolution of the exo-Jupiters, and hence
affect the evolution of their radii.

\item
For a Jupiter-like $B$-field strength, Ohmic dissipation is negligible
up to irradiation temperatures $T_{\rm irr} \la 1700-1800$~K, as far
as the power needed to influence the thermal evolution of the hot Jupiters.  As
the insolating flux becomes stronger, the magnitude of the dissipation
increases, and so does its penetration depth. The latter effect is
however much more pronounced in models without temperature inversions,
which have a hotter interior. Dissipation can occur down to levels of
several tens of bars, hence potentially affecting the radius evolution
of the exoplanet.  If Ohmic dissipation does play a major role in
radius inflation, a strong correlation with the strength of the
irradiating flux is naturally predicted. However, a large scatter is
also expected, due to the strong dependence of the Ohmic power on the
magnetic field strength, as well as on the atmospheric opacities.

\end{itemize}

\acknowledgements This work was partially supported by grant NSF
AST-1009396 (RP), the Zwicky Prize Fellowship (KH) and the STFC 
Advanced Fellowship (FP).  RP thanks the ETH for the wonderful hospitality
during the time that some of this work was carried out.  KH benefited from 
stimulating discussions conducted at the Exoclimes II conference, held at the 
Aspen Center for Physics.  We are grateful to 
Olivier Byrde and his team for supporting our use of the \texttt{Brutus} computing 
cluster at ETH Z\"{u}rich, which served as the workhorse for our simulations.
Finally, we thank the referee, Jonathan Fortney, for very useful comments
on the manuscript.

\appendix
\section{Zonal Wind, Temperature, Potential Temperature and Streamfunction Profiles}
\label{append}

\begin{figure}
\centering
\includegraphics[width=0.48\columnwidth]{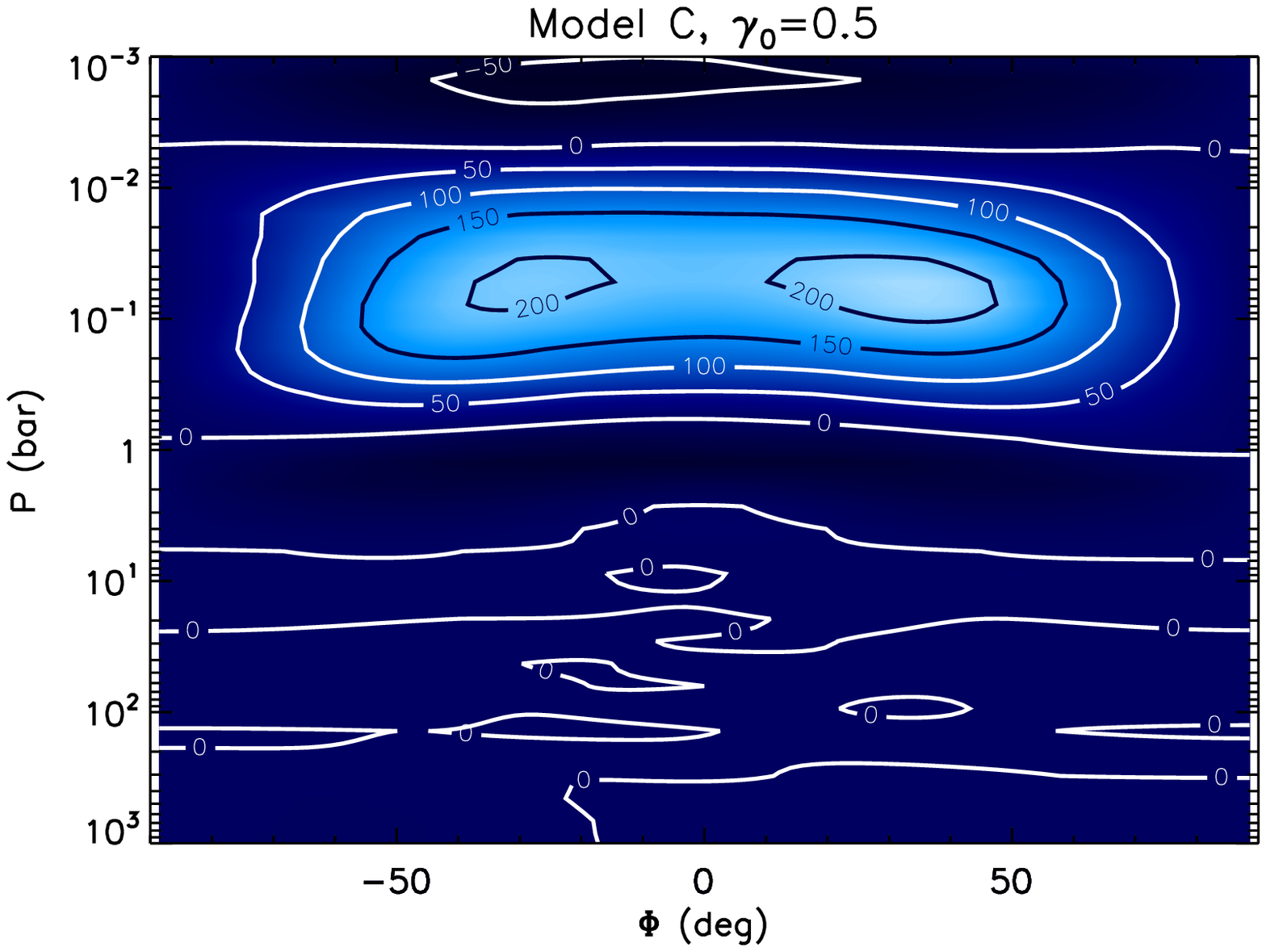}
\includegraphics[width=0.48\columnwidth]{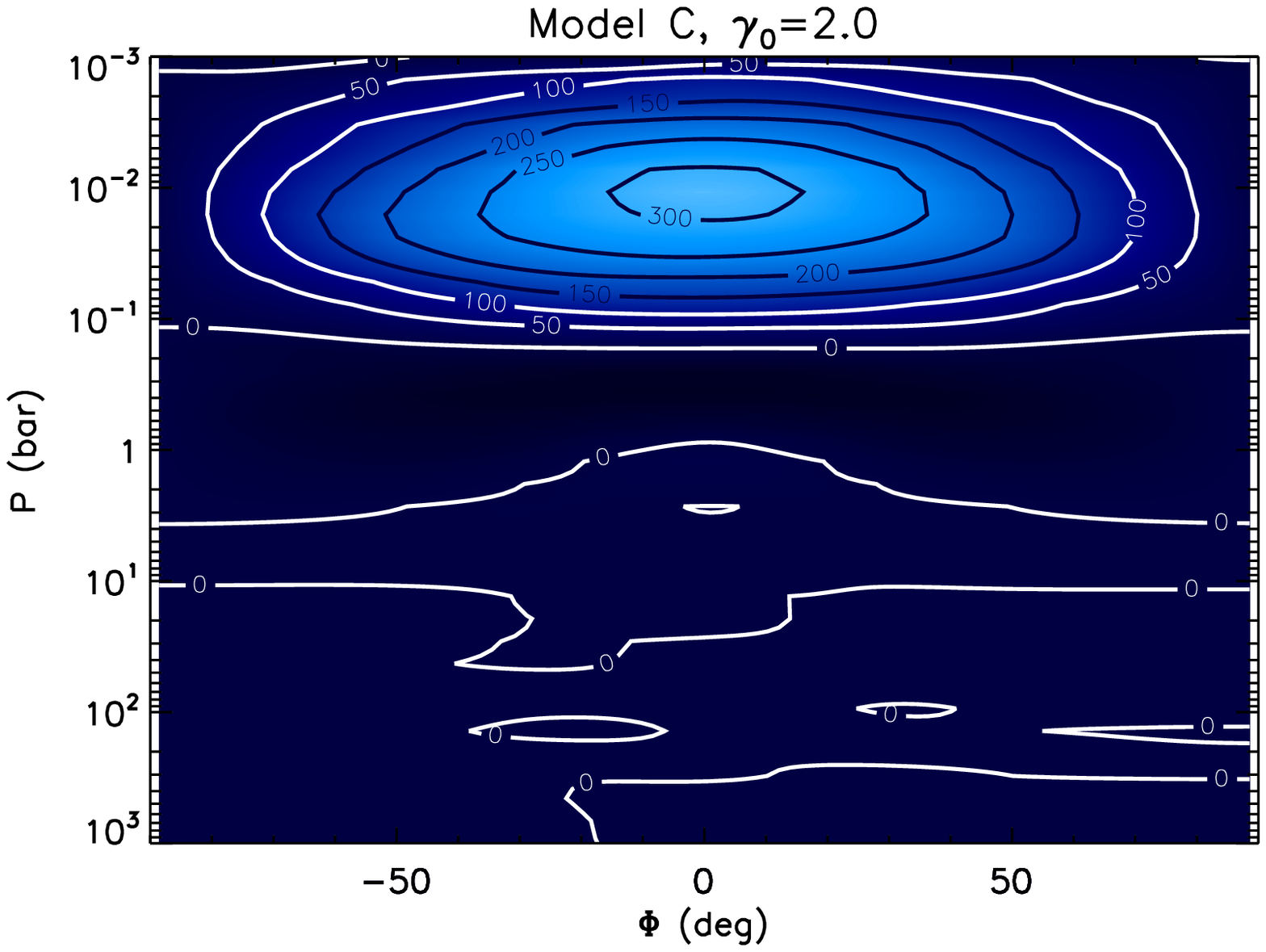}
\includegraphics[width=0.48\columnwidth]{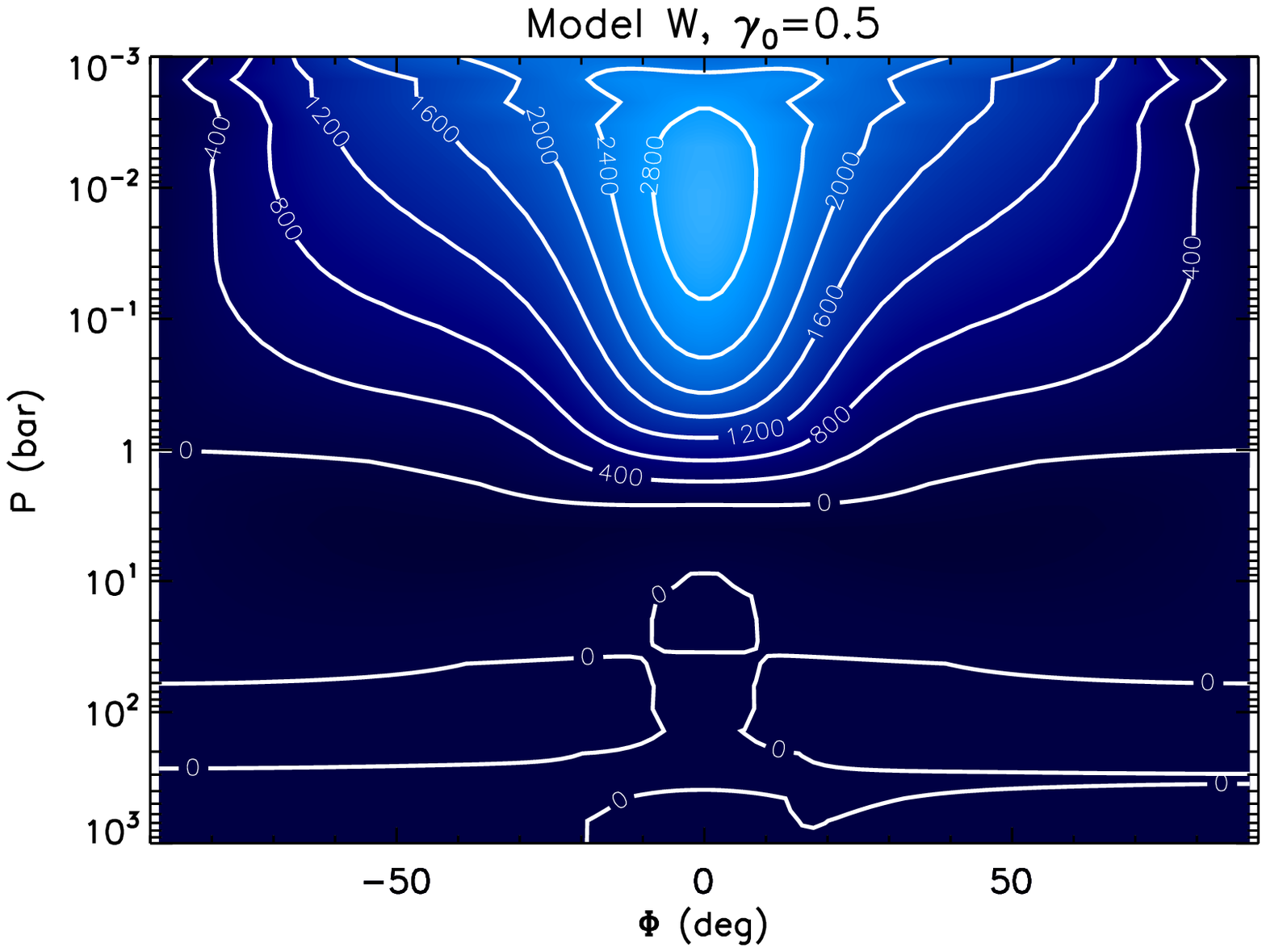}
\includegraphics[width=0.48\columnwidth]{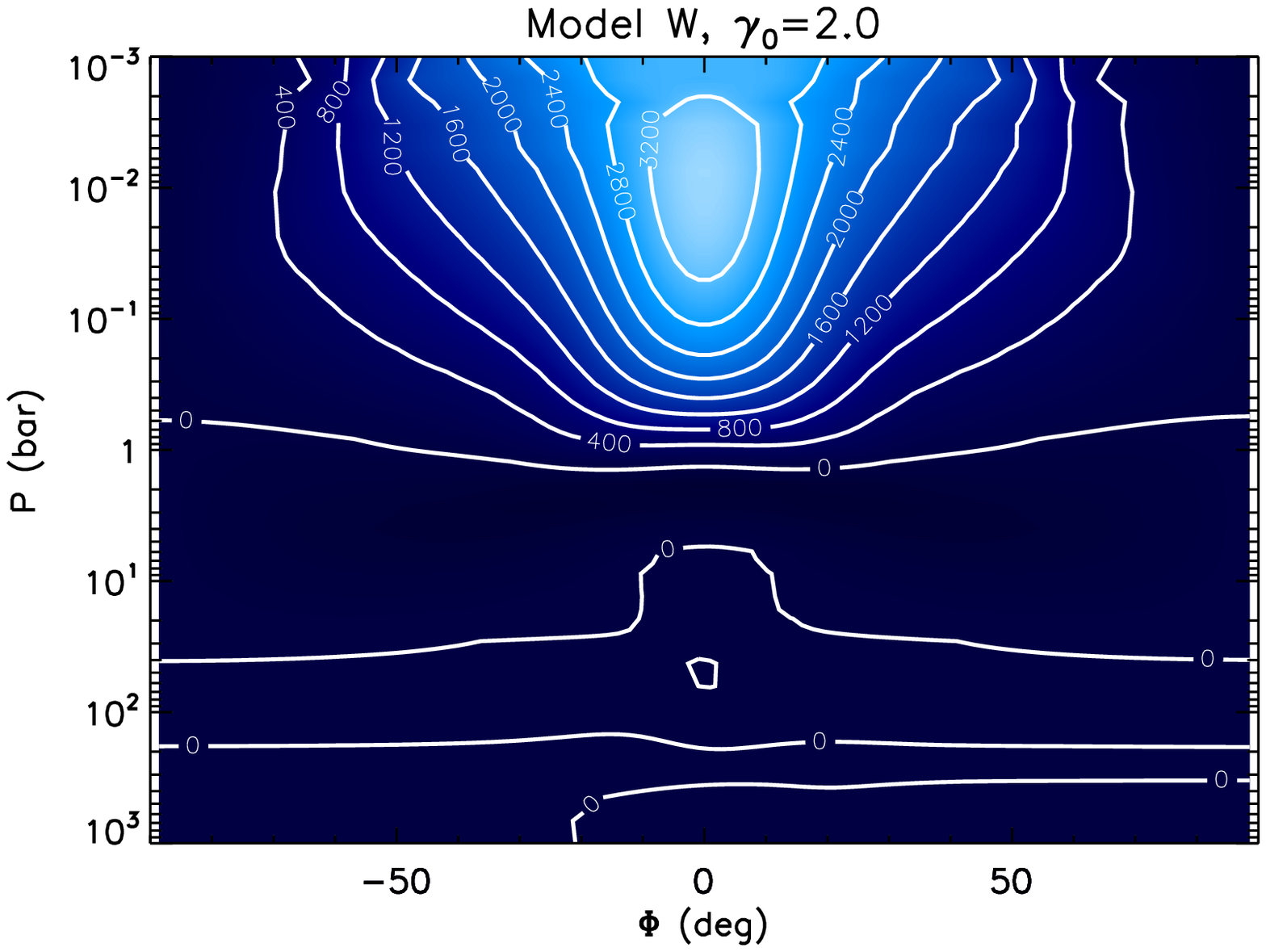}
\includegraphics[width=0.48\columnwidth]{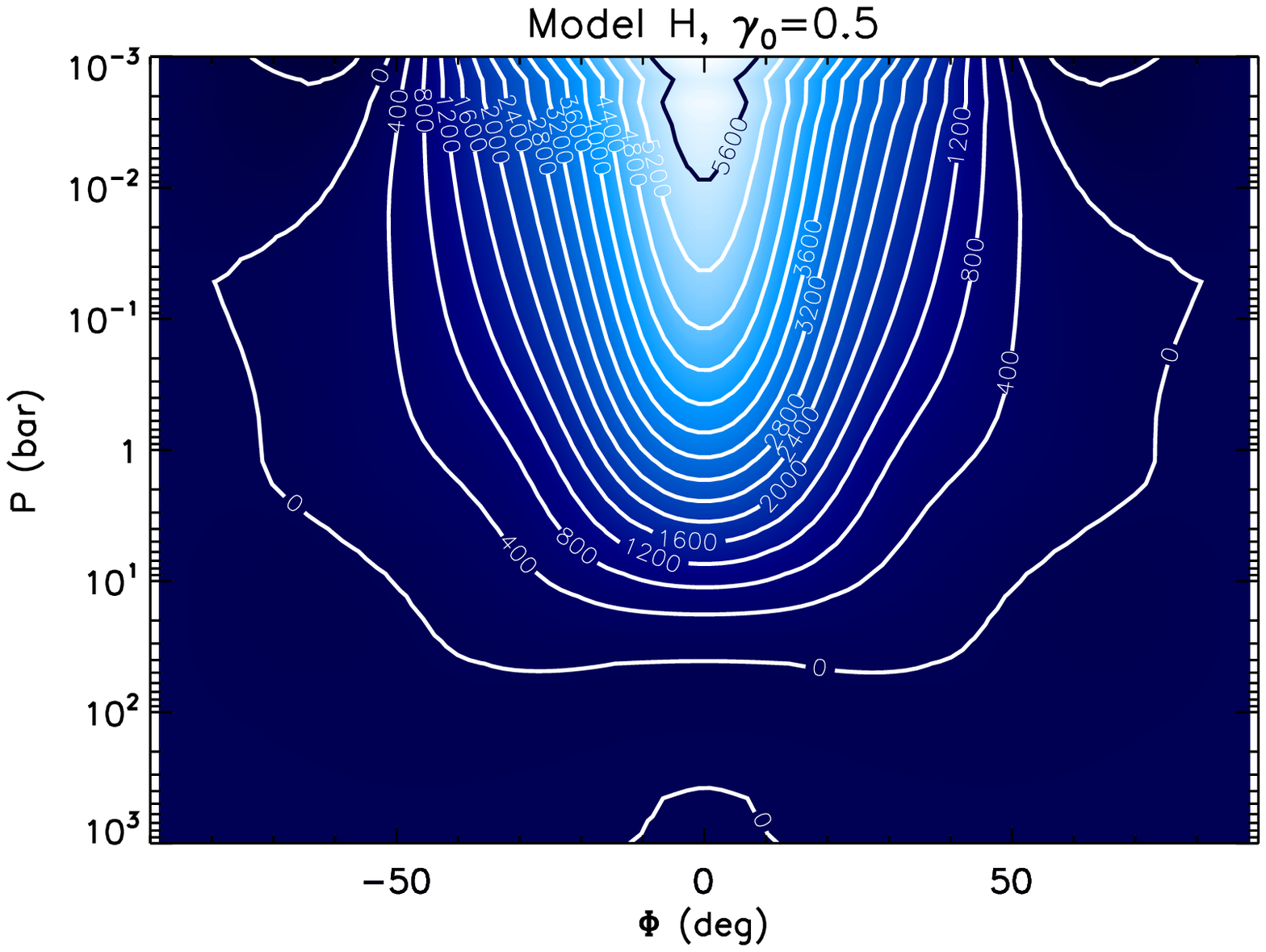}
\includegraphics[width=0.48\columnwidth]{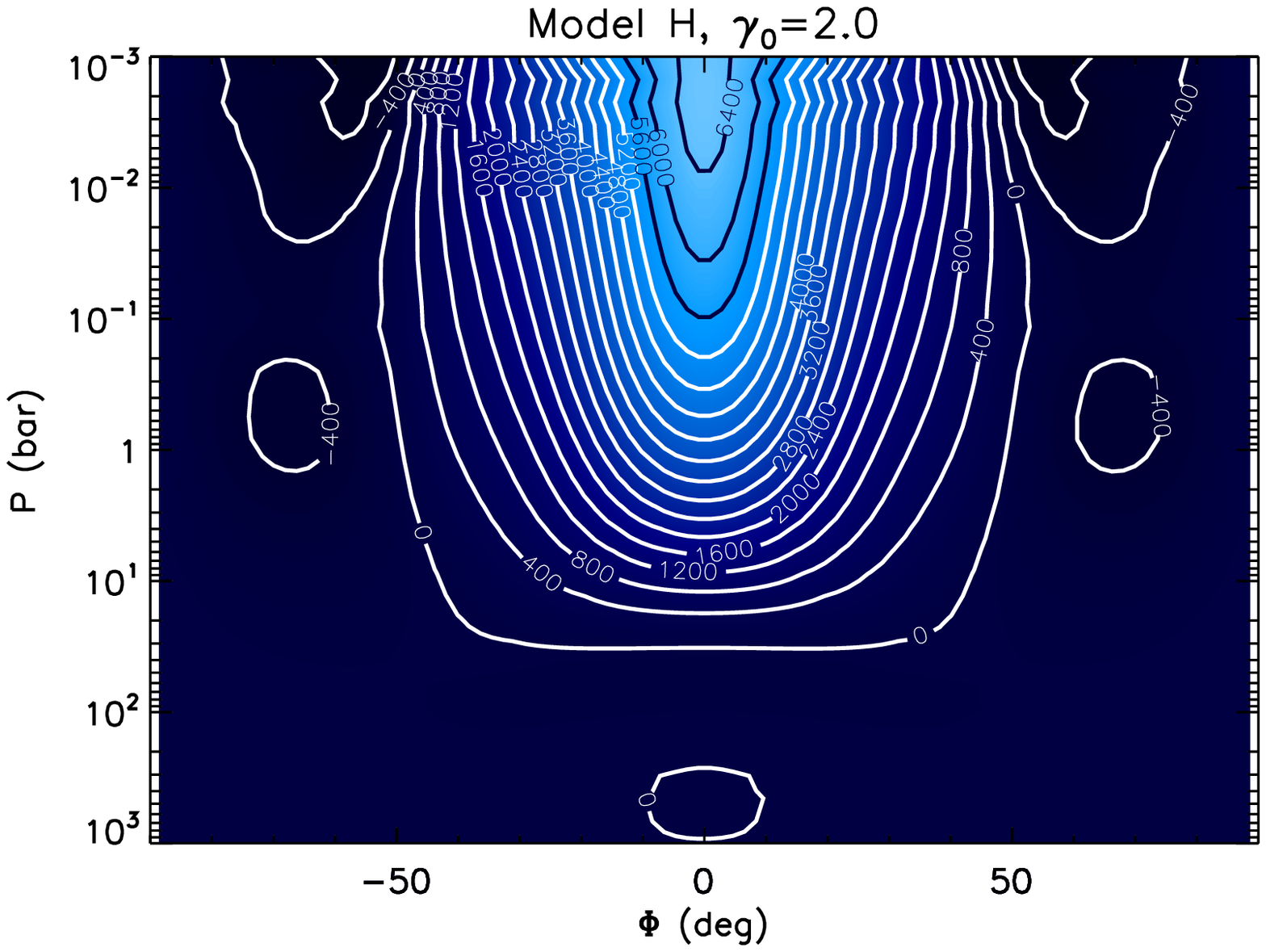}
\caption{Temporally- and zonally-averaged zonal wind profiles for Models C, W and H and for $\gamma_0=0.5$ and 2.  Contours are in units of m s$^{-1}$.}
\label{fig:zonal_wind}
\end{figure}

\begin{figure}
\centering
\includegraphics[width=0.48\columnwidth]{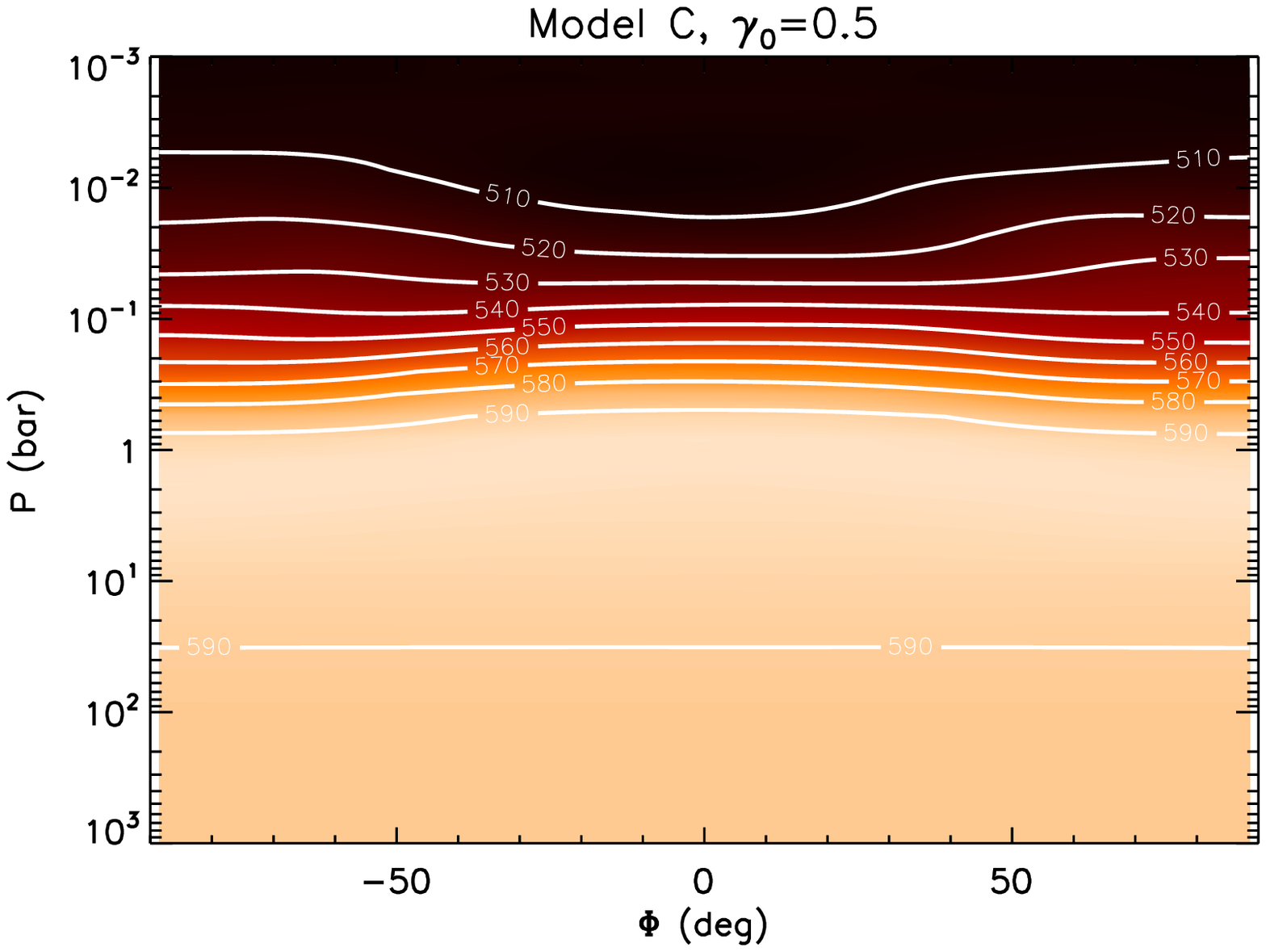}
\includegraphics[width=0.48\columnwidth]{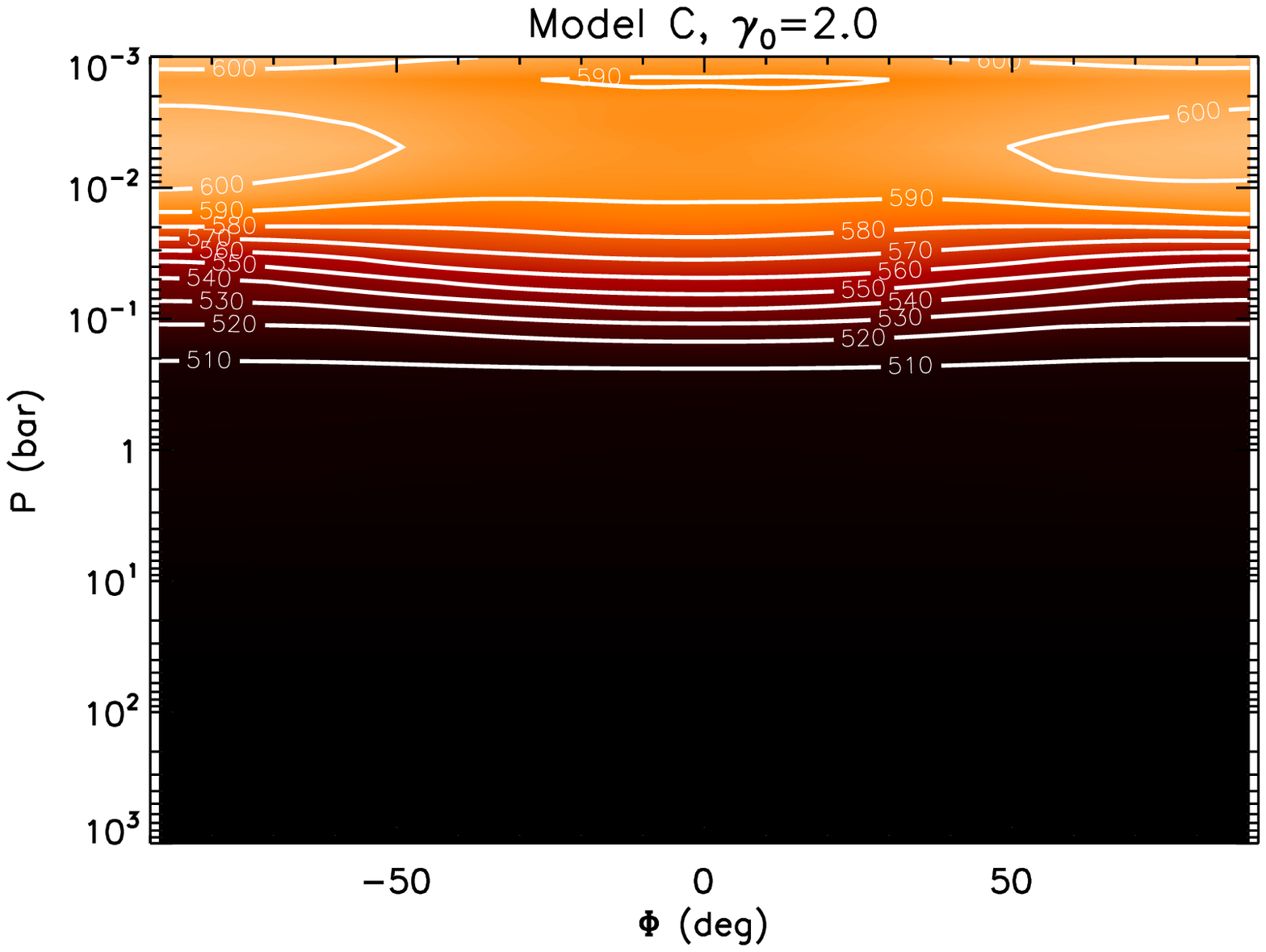}
\includegraphics[width=0.48\columnwidth]{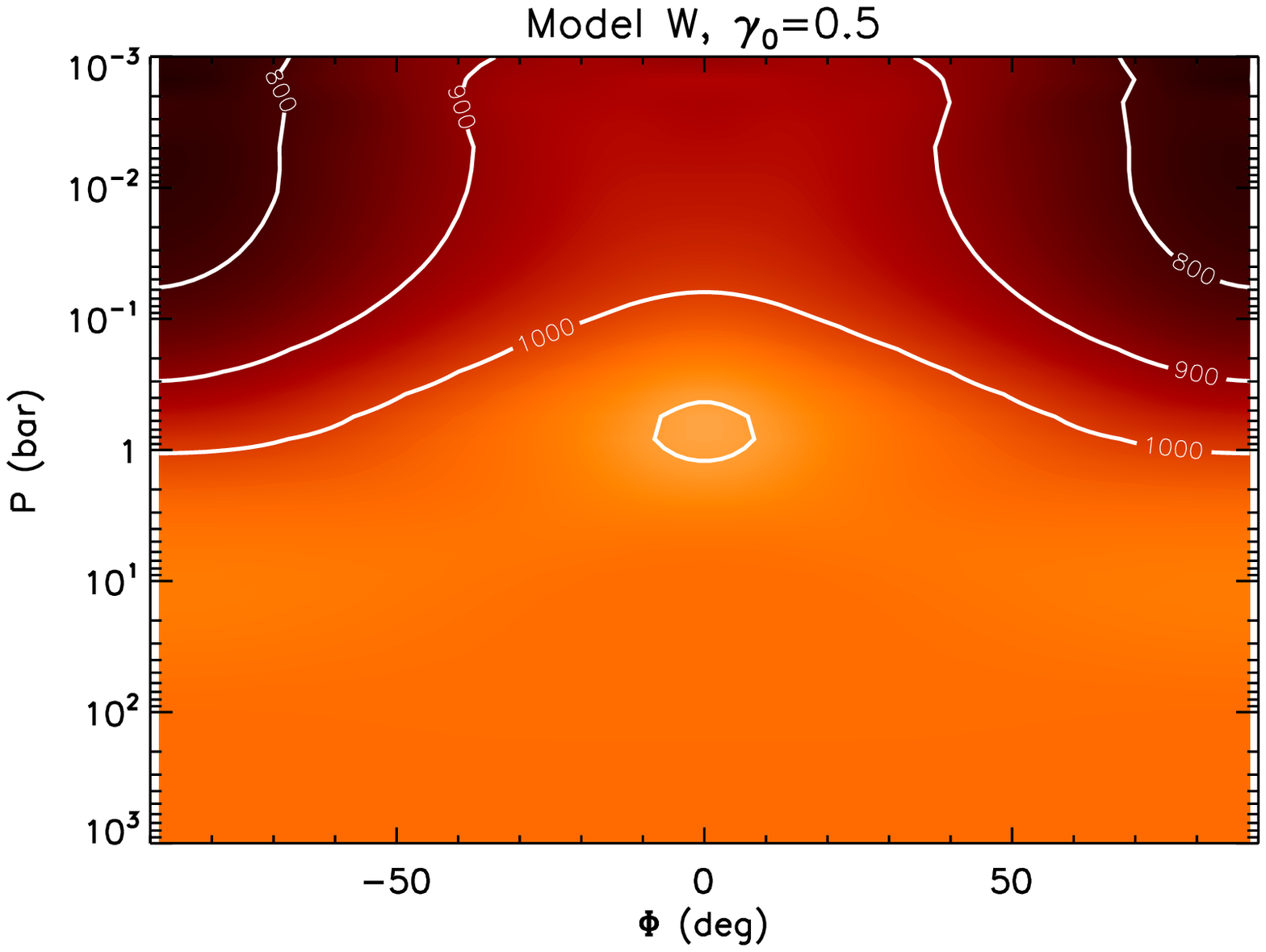}
\includegraphics[width=0.48\columnwidth]{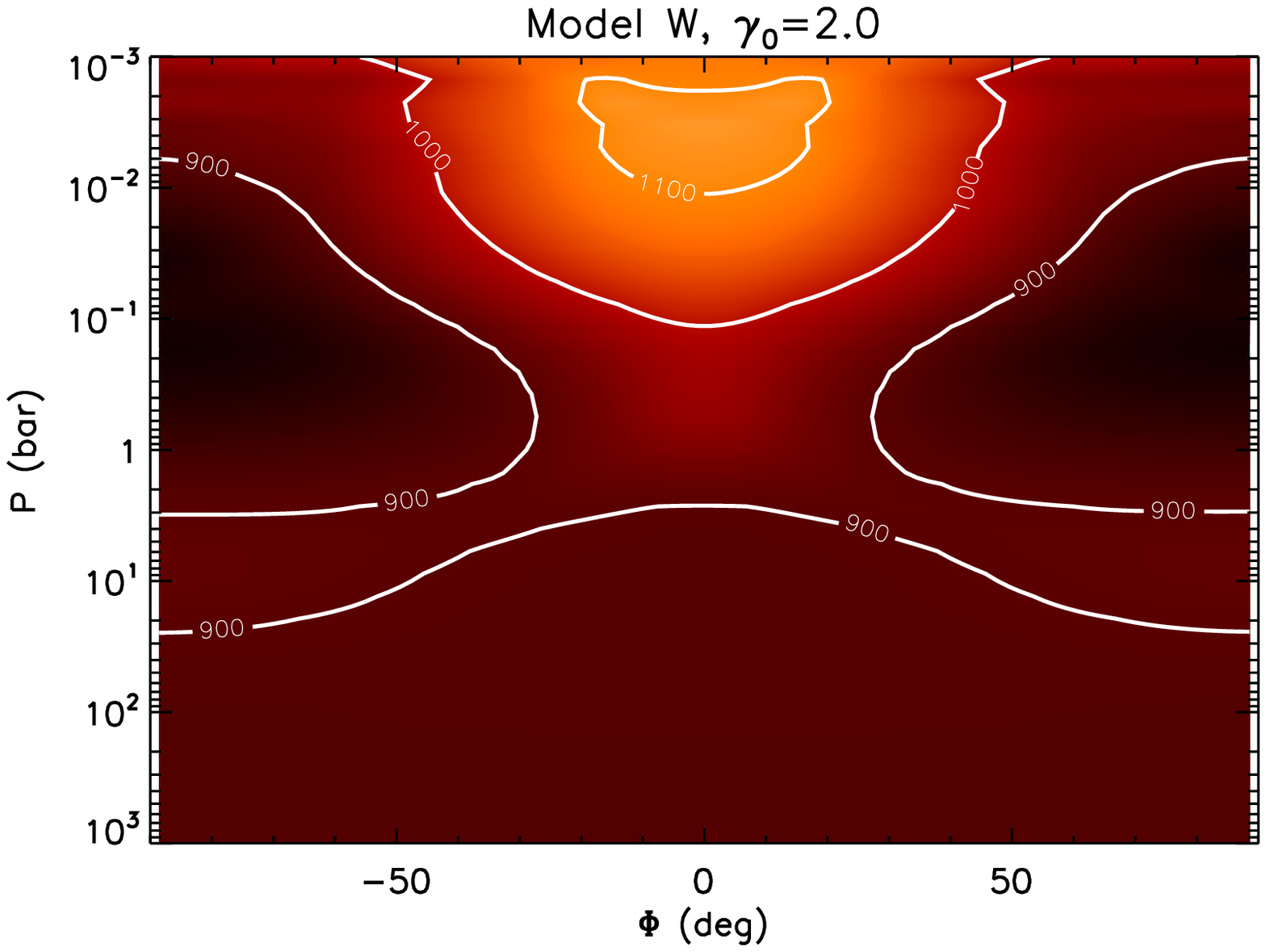}
\includegraphics[width=0.48\columnwidth]{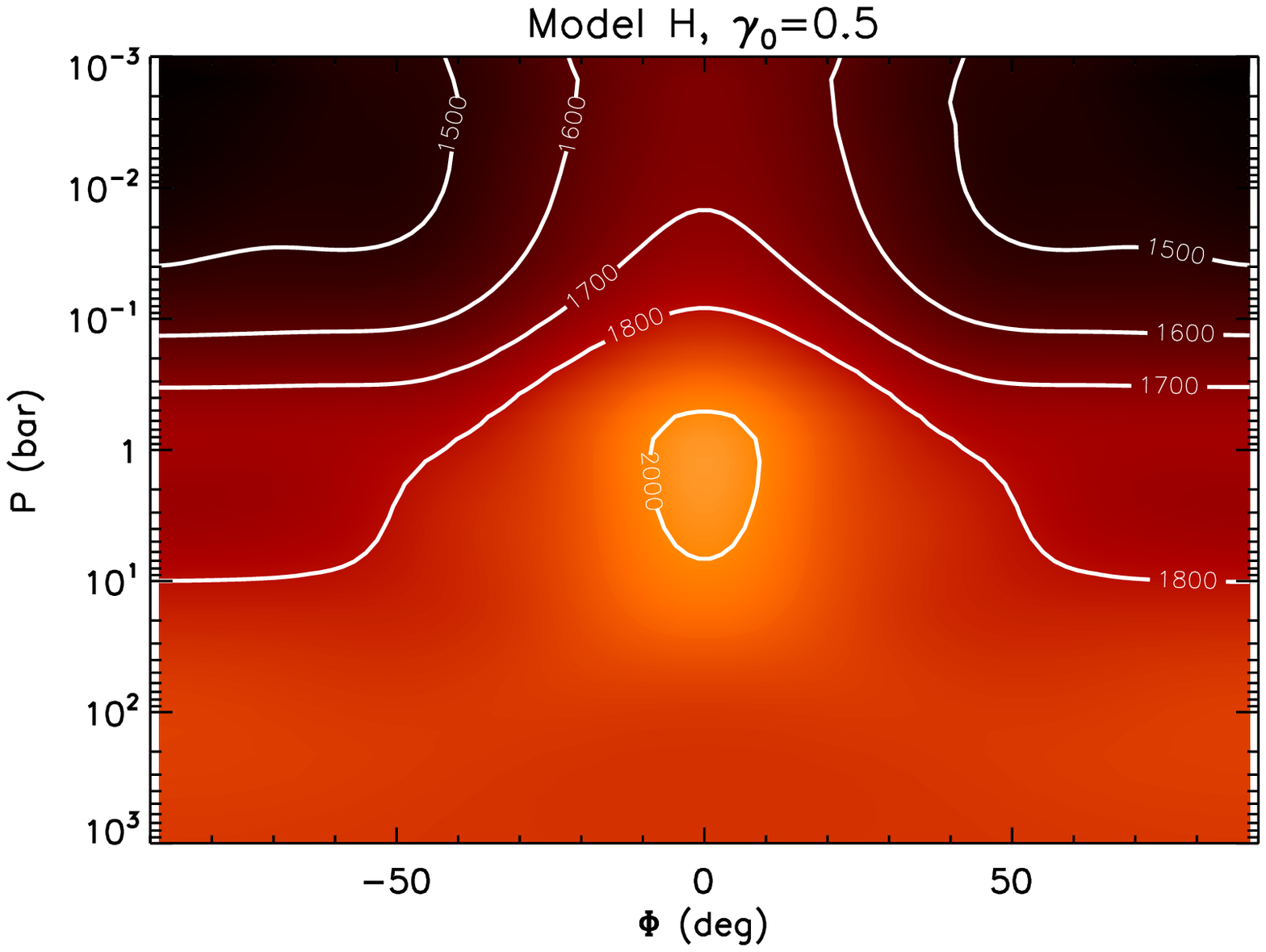}
\includegraphics[width=0.48\columnwidth]{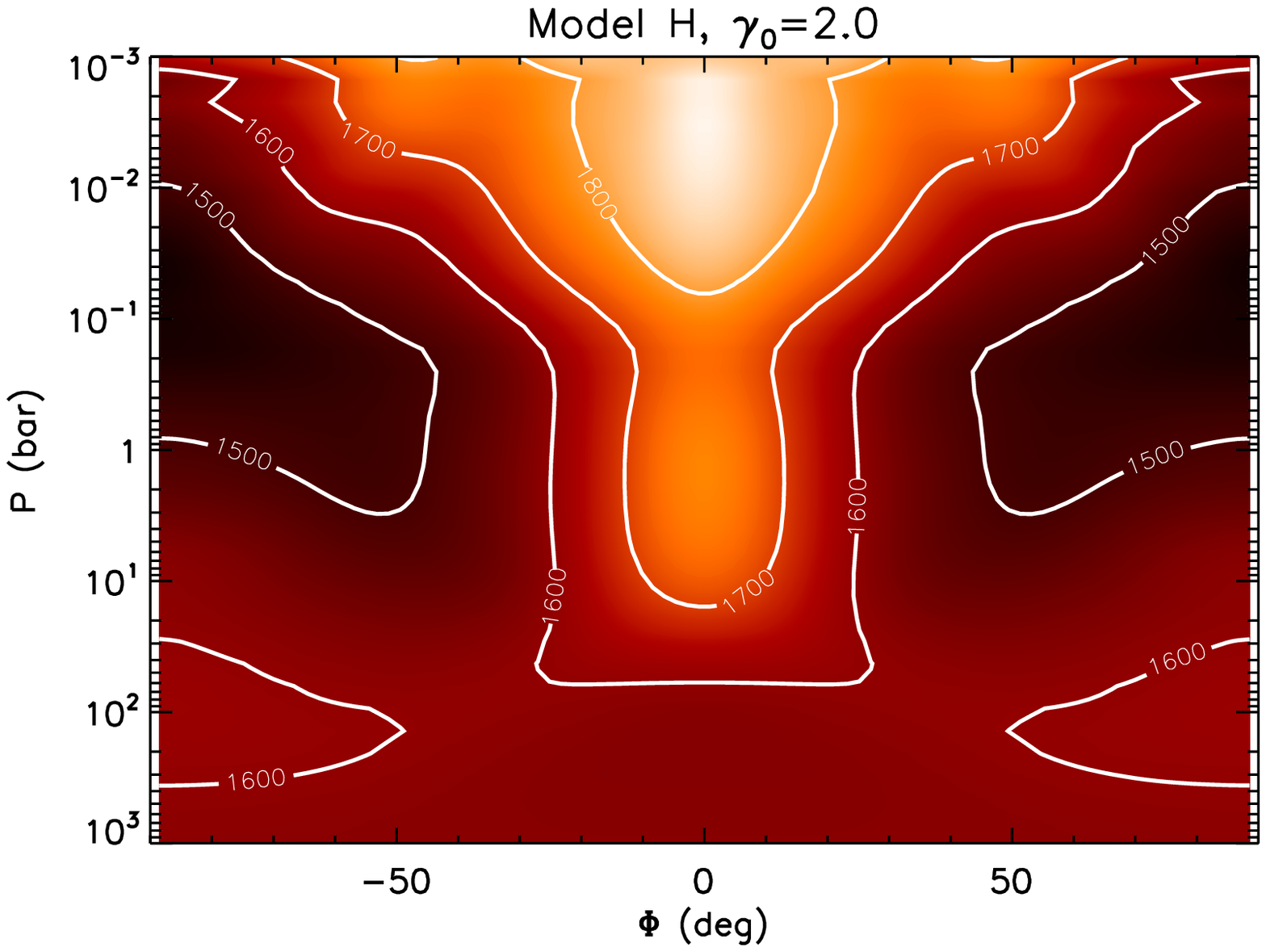}
\caption{Same as Figure \ref{fig:zonal_wind} but for the temporally- and zonally-averaged temperature profiles.  Contours are in units of K.}
\label{fig:zonal_temp}
\end{figure}

\begin{figure}
\centering
\includegraphics[width=0.48\columnwidth]{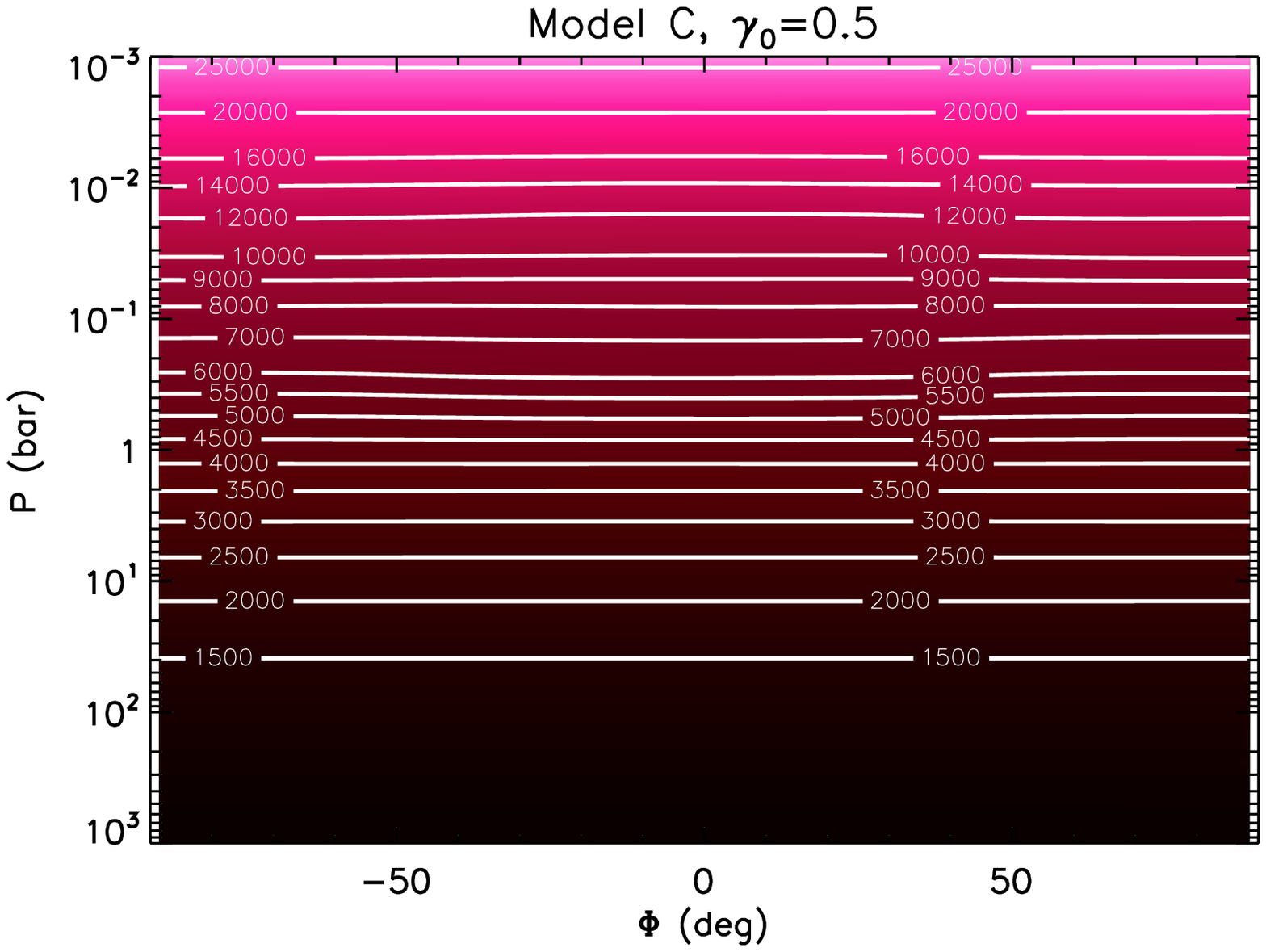}
\includegraphics[width=0.48\columnwidth]{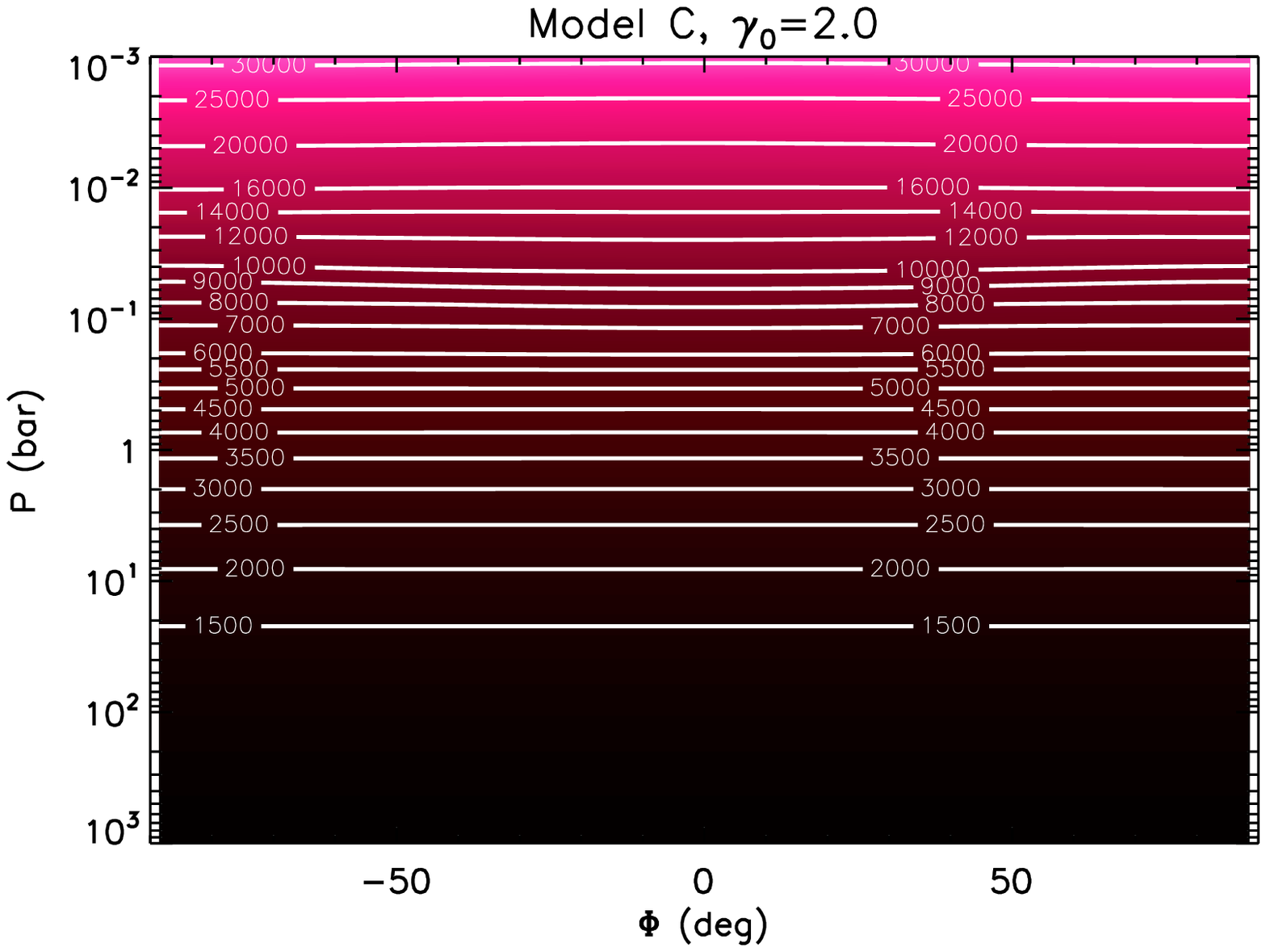}
\includegraphics[width=0.48\columnwidth]{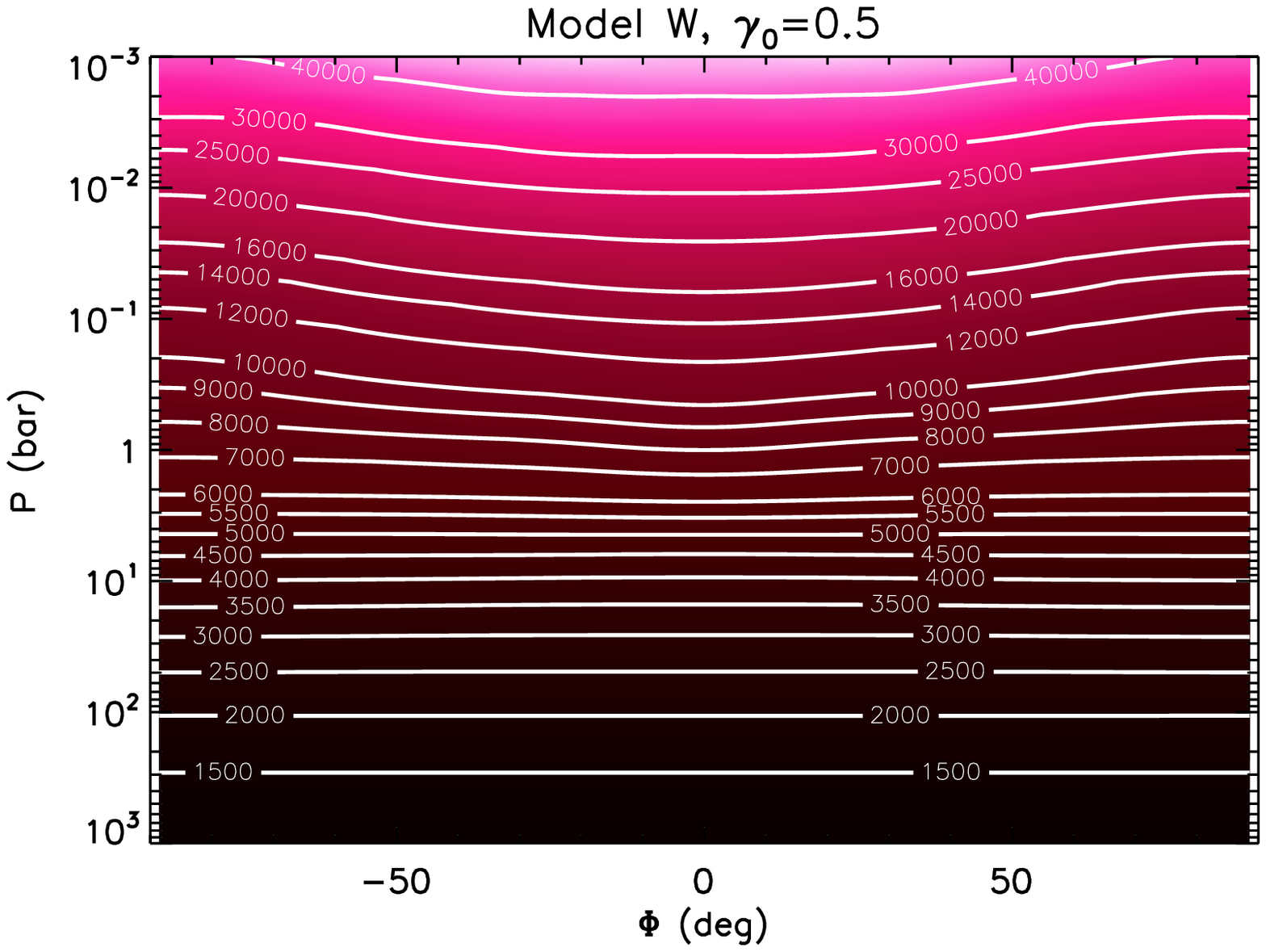}
\includegraphics[width=0.48\columnwidth]{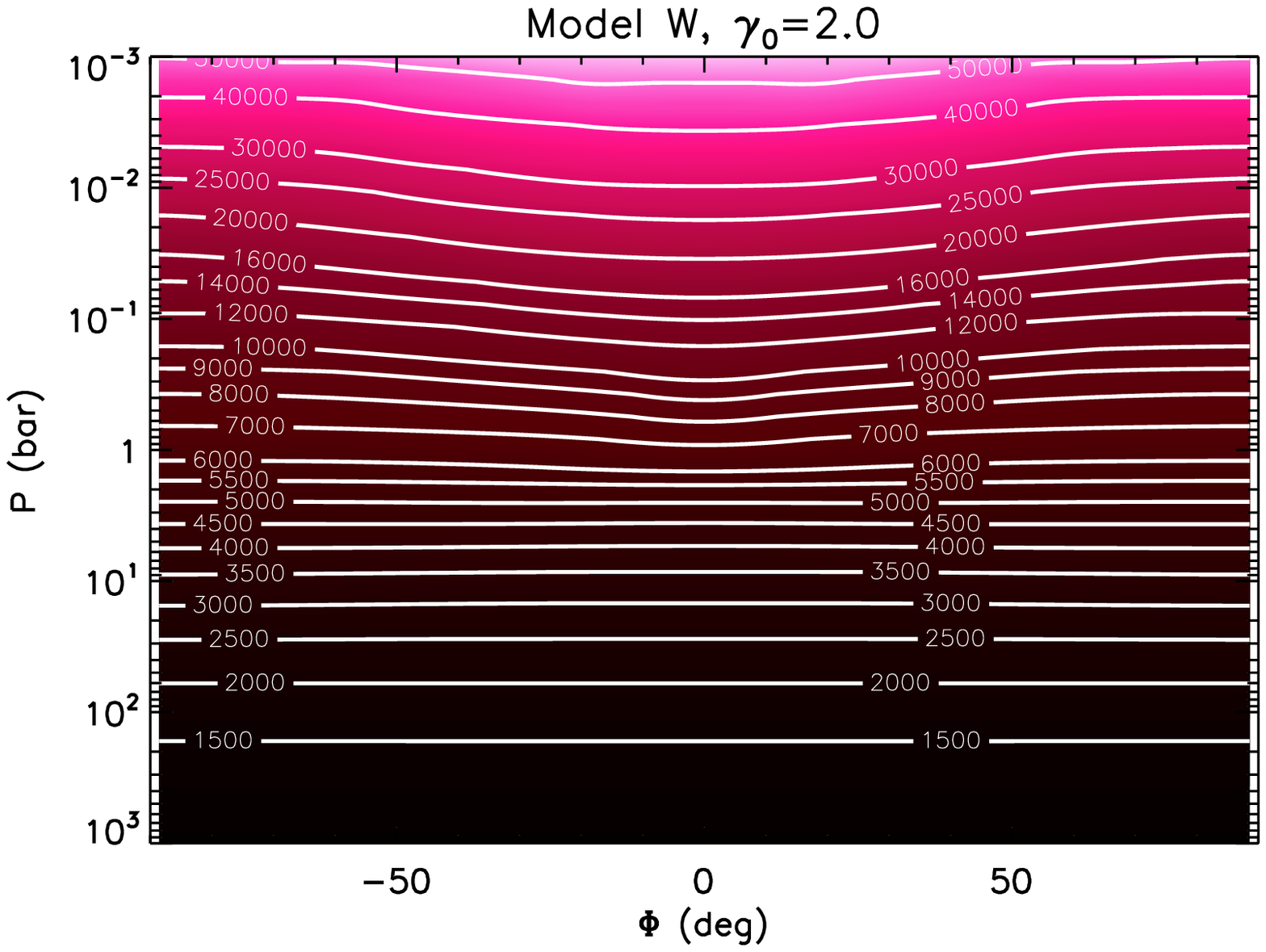}
\includegraphics[width=0.48\columnwidth]{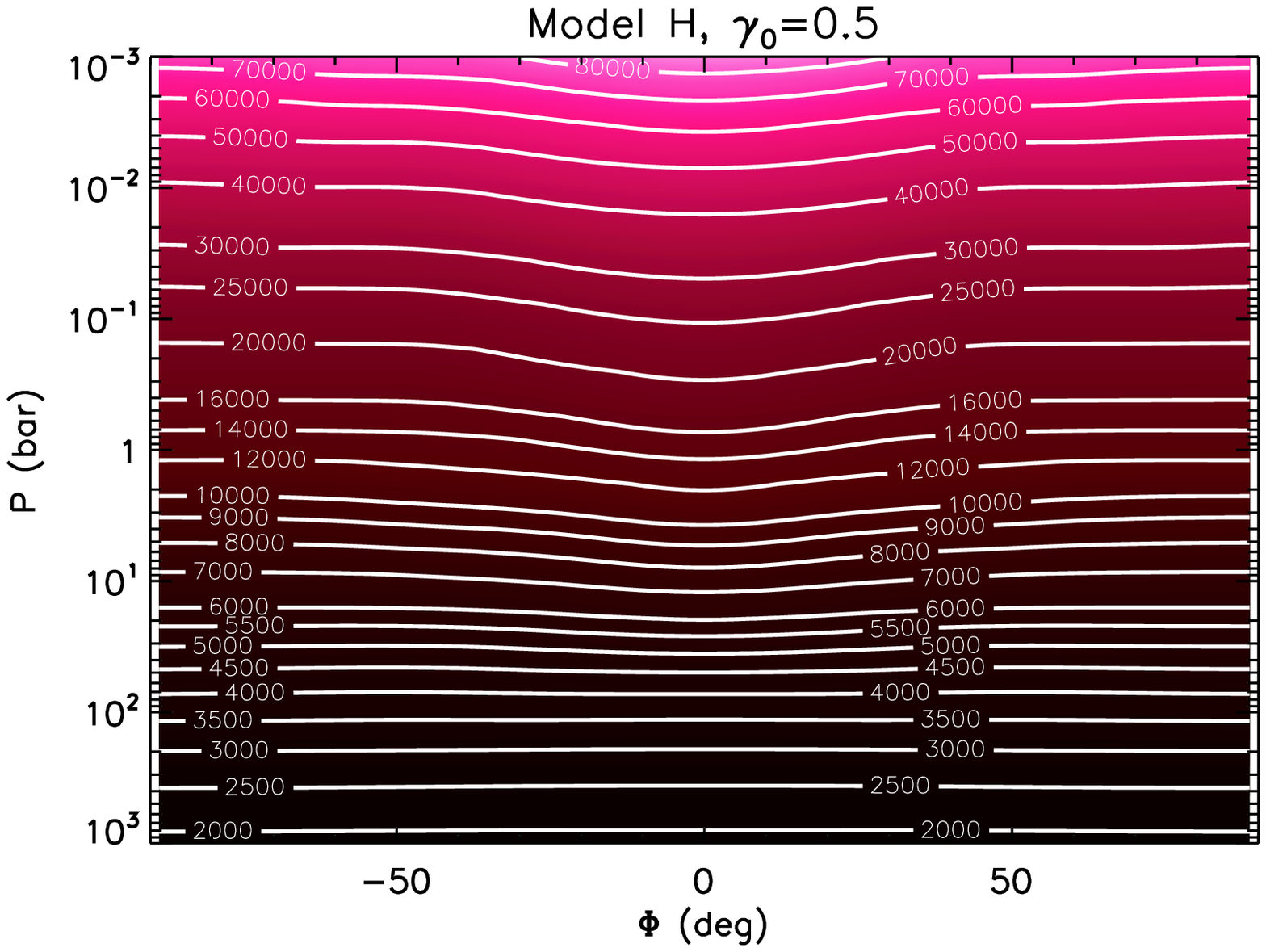}
\includegraphics[width=0.48\columnwidth]{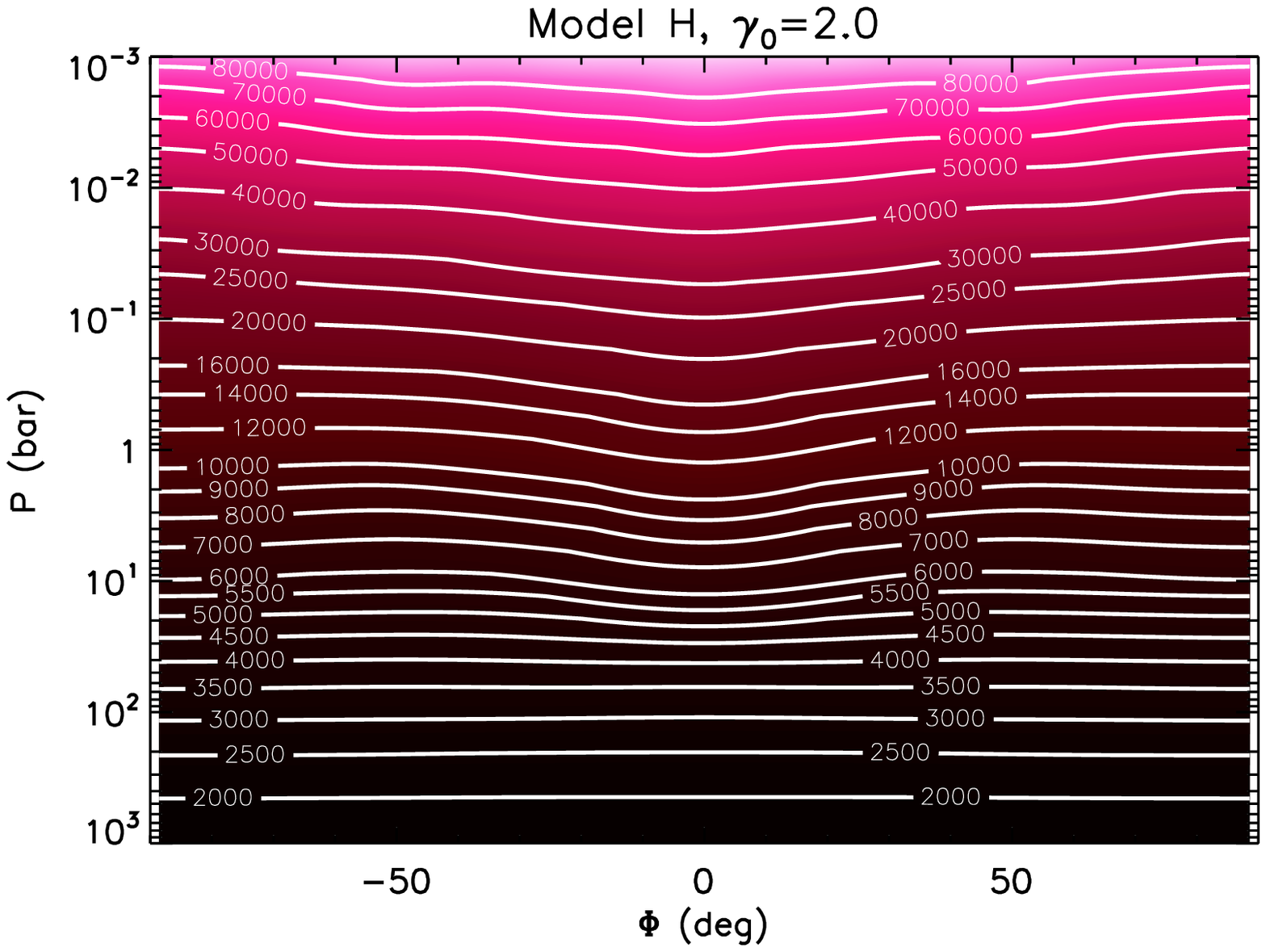}
\caption{Same as Figure \ref{fig:zonal_wind} but for the temporally- and zonally-averaged potential temperature profiles.  Contours are in units of K.  Lines of constant potential temperature are equivalent to lines of constant entropy (isentropes).}
\label{fig:zonal_potemp}
\end{figure}

\begin{figure}
\centering
\includegraphics[width=0.48\columnwidth]{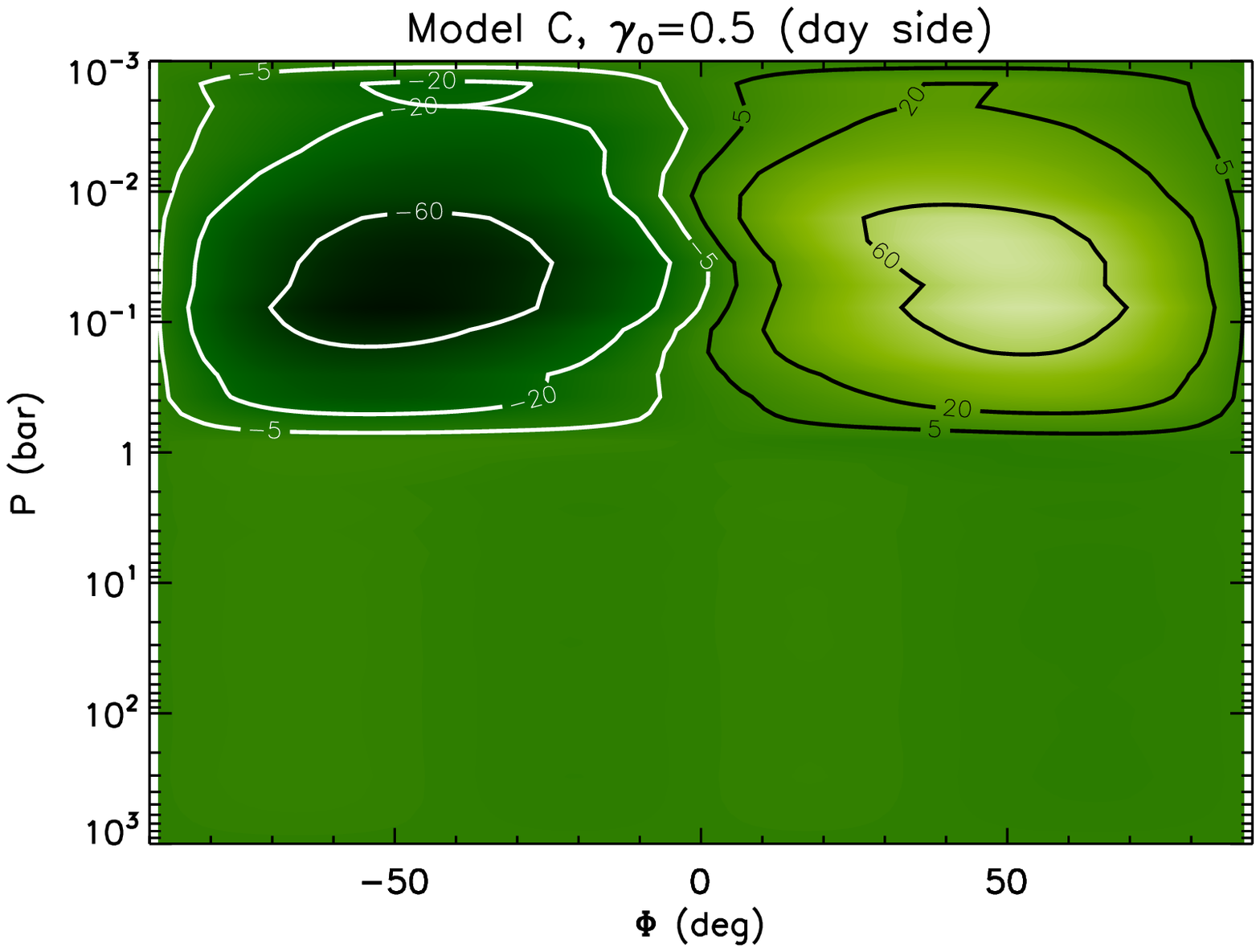}
\includegraphics[width=0.48\columnwidth]{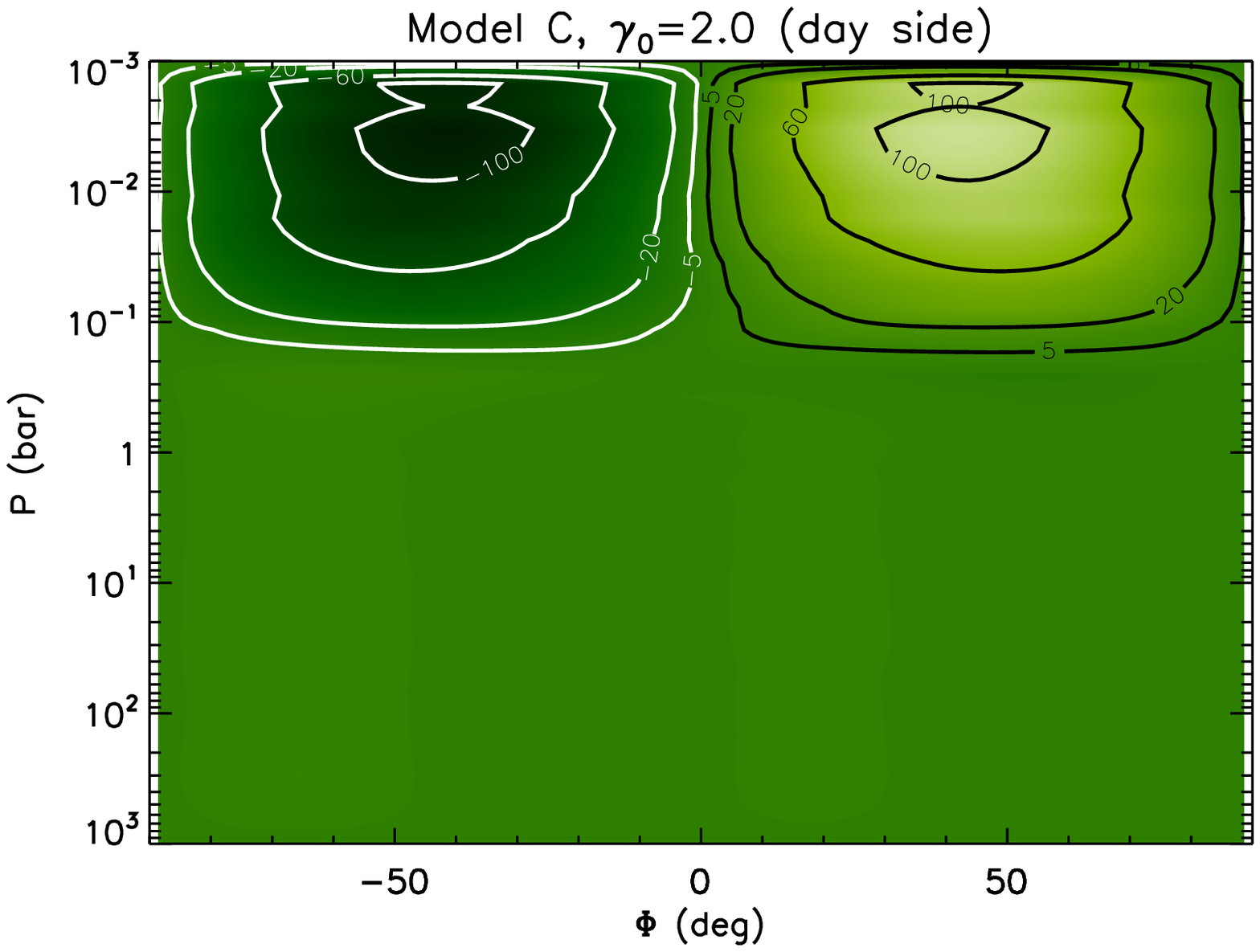}
\includegraphics[width=0.48\columnwidth]{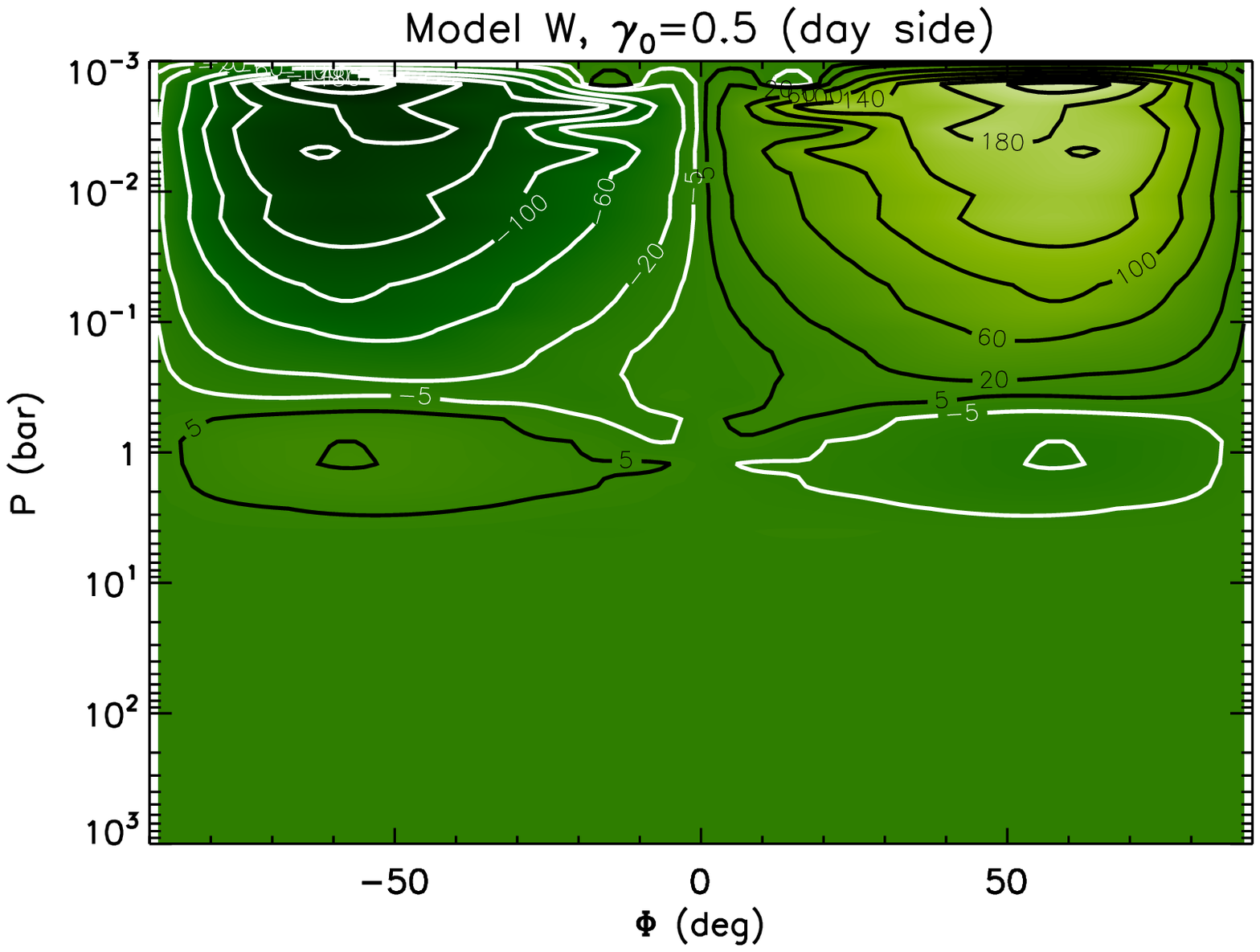}
\includegraphics[width=0.48\columnwidth]{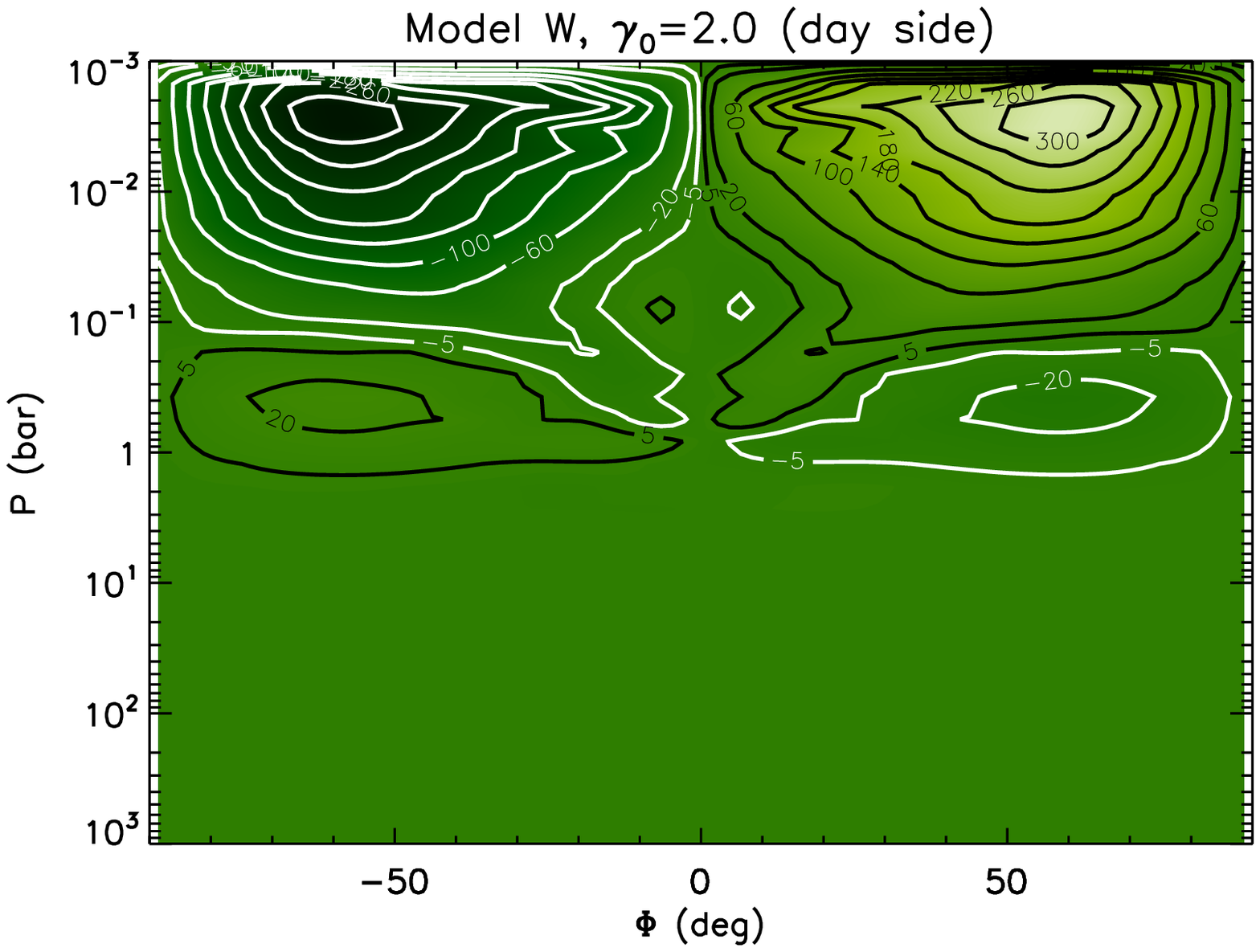}
\includegraphics[width=0.48\columnwidth]{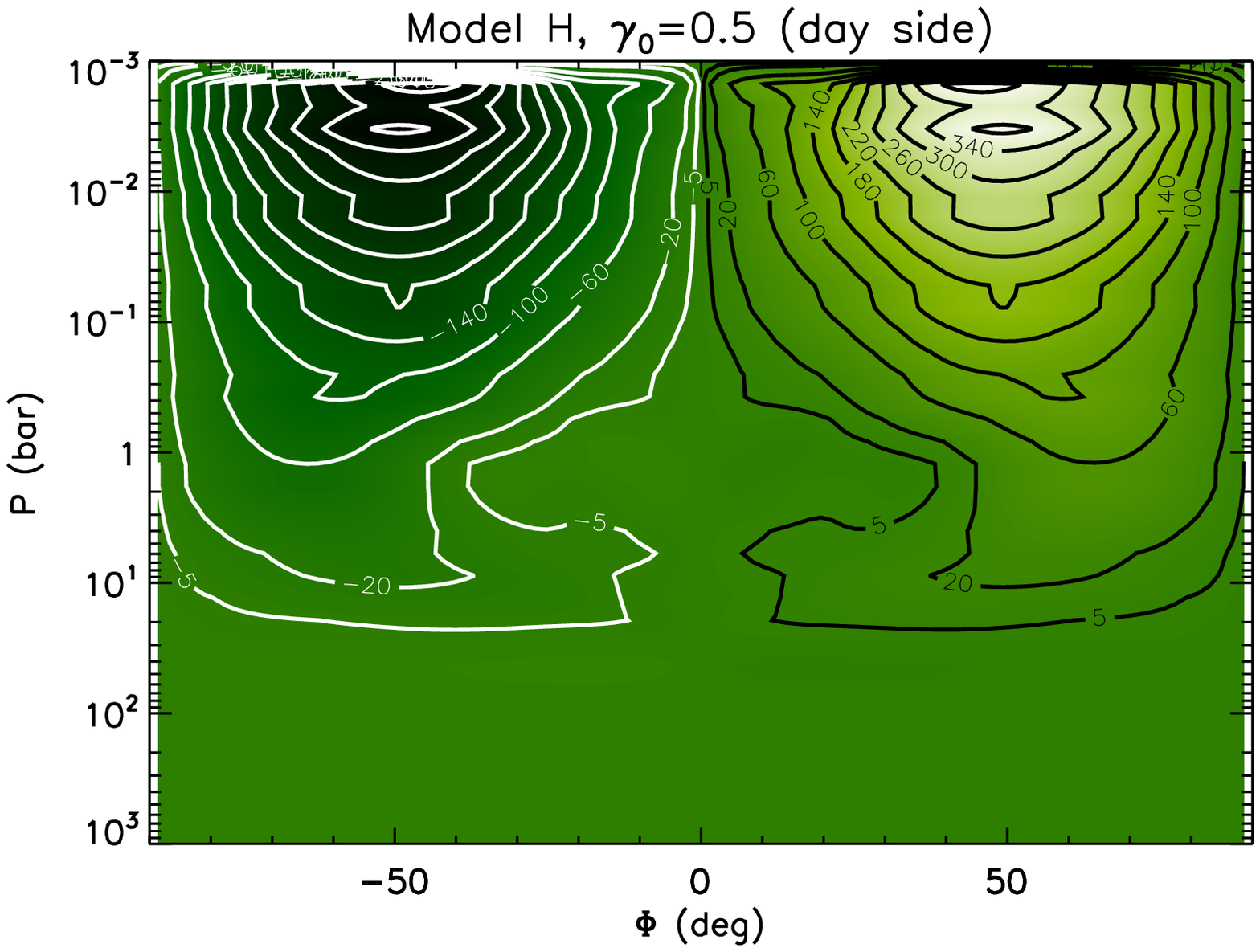}
\includegraphics[width=0.48\columnwidth]{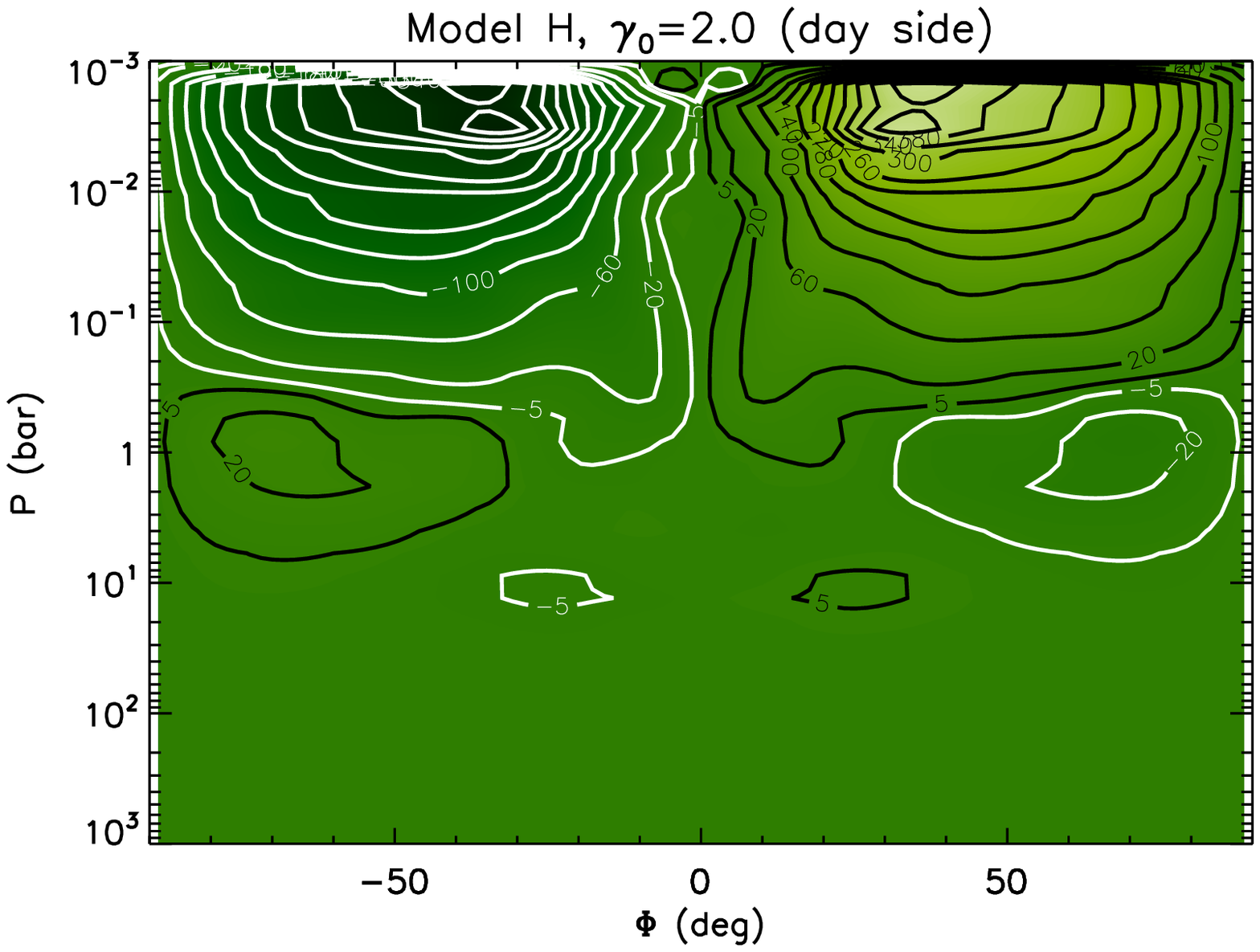}
\caption{Same as Figure \ref{fig:zonal_wind} but for the temporally- and zonally-averaged Eulerian streamfunction profiles only on the dayside hemisphere.  Contours are in units of $10^{13}$ kg s$^{-1}$.  Negative and positive values indicate clockwise and anti-clockwise circulation, respectively.}
\label{fig:streamfunction}
\end{figure}

For completeness, we include the temporally- and zonally-averaged zonal wind (Figure \ref{fig:zonal_wind}), temperature (Figure \ref{fig:zonal_temp}), potential temperature (Figure \ref{fig:zonal_potemp}) and Eulerian mean streamfunction (Figure \ref{fig:streamfunction}) profiles as functions of latitude and pressure/height.  The contours of constant potential temperature are equivalent to contours of constant specific entropy.  As already noted in \cite{hfp11}, computing the Eulerian mean streamfunction only on the dayside hemisphere strictly violates the definition of a streamfunction, but it does illustrate the nature of the global circulation pattern and the depth to which it penetrates.

\label{lastpage}

\end{document}